\documentclass{optica-article}
\journal{opticajournal}
\articletype{Research Article}

\usepackage{array} 

\usepackage{accents}
\usepackage{bm}

\usepackage{appendix}
\usepackage{caption}
\usepackage{subcaption}
\usepackage{tabularx}

\usepackage[normalem]{ulem}

\begin{document}

\title{Correcting astigmatism and ellipticity in Gaussian beams using a cylindrical lens pair with tunable focal lengths}



\author{Soroush Khoubyarian,\authormark{1, 2} Anastasiia Mashko\authormark{1, 2}, and Alexandre Cooper\authormark{1, 2, *}}

\address{
\authormark{1}~Department of Physics and Astronomy, University of Waterloo, 200 University West Avenue, Waterloo, ON N2L 3G1, Canada

\authormark{2}Institute for Quantum Computing, University of Waterloo, 200 University West Avenue, Waterloo, ON N2L 0A4, Canada
}

 \email{\authormark{*}alexandre.cooper@uwaterloo.ca} 


\begin{abstract*}
Correcting astigmatism and ellipticity in laser beams is critical for improving performance in many applications like microscopy, atomic physics, quantum information processing, and advanced manufacturing. Passive correction methods based on cylindrical lens telescopes require choosing lenses with precise focal lengths, effectively limiting the range of tunability when using standard catalog optics. Active solutions based on diffractive optical elements can achieve superior performance, but they are bulky, expensive, and suffer from finite diffraction efficiency and added complexity. Here, we introduce a simple method to convert astigmatic elliptical beams into circular Gaussian beams without astigmatism. Our method comprises three cylindrical lenses. The first lens focuses the beam along its major axis to create a plane where the intensity profile is radially symmetric. The second and third lenses are placed one behind the other in that plane at a relative angle, acting as a biaxial lens pair with tunable focal lengths. By adjusting the relative angle of the lenses, the two separate beam waists of the astigmatic beam can be overlapped, resulting in a circular Gaussian beam without astigmatism. We theoretically validate our method, numerically quantify its robustness to experimental imperfections, and experimentally demonstrate its ability to circularize the output beam of a commercial laser source. Our method corrects astigmatism and ellipticity in laser beams without requiring precise focal length matching, offering greater flexibility than other passive solutions and greater cost-effectiveness than active methods. Its simple and compact design makes it well suited for integration into both tabletop optical setups and industrial systems.
\end{abstract*}

\section{INTRODUCTION}
Correcting aberrations and imperfections in laser beams emitted by commercial laser sources is a common problem in laser optics~\cite{Hanna1969, Clarkson1996,  Knecht2007, Serkan2008, Tian2013, Sun2015}. This problem is typically encountered when operating free-space-emitting diode lasers, which produce elliptical and astigmatic beams~\cite{Serkan2008, Hasan2016}. It is also encountered when operating titanium:sapphire lasers, where astigmatism results from the Brewster-angle cut of the gain crystal and the potential alignment errors within the optical cavity~\cite{Hanna1969, Ramirez2016, Spectraphysics3900s}. Addressing this problem is critical in many applications, such as those involving high-power optical systems where fiber coupling and spatial filtering are not viable solutions.

\begin{figure}[t!]
\centering
\includegraphics[]{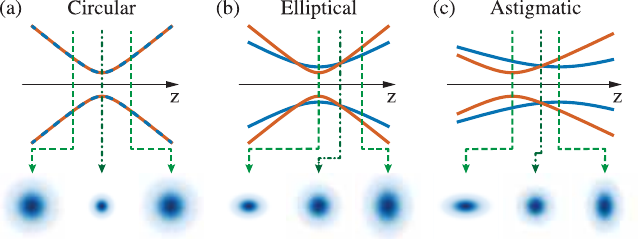}
\caption{
\label{fig:gaussian_beams}
\textbf{Gaussian beams.}
(a)~A circular Gaussian beam has a circular intensity profile with radial symmetry.
(b)~An elliptical Gaussian beam has distinct divergence angles along its major (blue) and minor (red) axes. The beam radius along each axis is minimized in the same focal plane, where the intensity profile is elliptical. The intensity profile becomes circular in two planes symmetrically located on either side of the beam waist.
(c)~An astigmatic Gaussian beam is an elliptical beam whose beam waists do not overlap. There exists at least one plane where the intensity profile is circular.
}
\end{figure}

\begin{figure}[t!]
\centering
\includegraphics[]{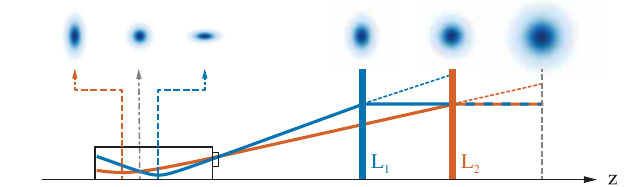}
\caption{
\label{fig:two_lenses}
\textbf{Typical two-cylindrical-lens solution.} An astigmatic elliptical Gaussian beam can be converted into a circular Gaussian beam by placing two cylindrical lenses, $L_1$ and $L_2$, each oriented along one of the principal axes of the diverging beam. Each cylindrical lens is placed approximately one focal length away from its corresponding beam waist.
}
\end{figure}

A common approach to circularizing an elliptical beam (see Fig.~\ref{fig:gaussian_beams}) involves two cylindrical lenses oriented along the principal axes of the beam (see Fig.~\ref{fig:two_lenses}). This approach benefits from low cost and simplicity; however, it requires precisely choosing the focal lengths of the lenses according to a precise ratio, which is not always possible when sourcing lenses from standard optics catalogs. These lenses may also deviate from being perfectly uniaxial, leading to unwanted changes in the beam divergence along the weak axis of the cylindrical lens, thus preventing circularization. Another approach to circularizing the beam emitted by diode lasers involves using anamorphic prism pairs combined with a circular clipping aperture. This method benefits from a compact and tunable design, low optical loss, high-power handling, and intrinsic correction of chromatic dispersion. However, these prism pairs correct only for the ellipticity of the beam far away from the beam waist by expanding or compressing the beam in one direction; they do not affect the relative separation between the beam waists, and thus do not correct astigmatism~\cite{Murty1984, Sun2015}.
Other passive circularization methods exist and have demonstrated high circularity and astigmatism correction (see e.g., Refs.~\cite{Xiao2000, Yang2008, Tian2013, Serkan2009}); however, these approaches provide limited tunability and depend on specialized, custom-made components tailored to the properties of the input beam.

Active methods based on active optical diffracting elements, such as spatial-light-modulators, benefit from a high level of tunability, enabling the active suppression of high-order aberration terms~\cite{Cheng2020, Bowman2010}; however, these elements are bulky, expensive, and rely on complex software optimization to achieve optimal performance. They also suffer from power losses due to their finite diffraction efficiency, making them unsuitable for compact, energy-efficient, portable systems. Their cost and complexity are typically not justified for simple, compact applications such as maximizing the coupling efficiency of light in an optical fiber~\cite{Wagner1982}.

In this paper, we introduce a simple method to correct astigmatism and ellipticity in laser beams. Our solution employs three standard cylindrical lenses mounted on rotation mounts~(see Fig.~\ref{fig:three_lenses}). The first cylindrical lens enables the creation of a plane in which the intensity profile is circular. The other two cylindrical lenses are placed one behind the other in that plane at a relative angle, acting as an effective biaxial lens with tunable focal lengths that can match the divergence angles along the two principal axes of the beam. Our method reduces alignment constraints by eliminating the need for precise selection of the focal lengths of these lenses. It is both compact and flexible, allowing for the use of off-the-shelf components with a broad range of focal lengths. Our method remains effective even when the cylindrical lenses are not perfectly uniaxial, accommodating manufacturing variations without significant loss of performance. Our solution can be used to circularize laser beams in various settings such as neutral-atom quantum computing, where high-power, circular Gaussian beams are critical. For example, these circular beams enable trapping individual neutral atoms in large arrays of optical traps~\cite{Bowman2010, Chew2024, Wulff2006}, displacing atoms to correct for lost atoms in large arrays of optical tweezers~\cite{Cimring2023, ElSabeh2023, Afiouni2025, 
Dadpour2025}, and performing site-selective quantum gates with minimal crosstalk~\cite{BinaiMotlagh2023, Lim2025}. A similar solution has been used to construct an anamorphic beam expander for adjusting the magnification and aspect ratio of a laser beam at a fixed plane~\cite{DeHoog2021}. It is also employed in the Jackson Cross Cylinder, which uses two cylindrical lenses with equal but opposite focal lengths mounted one behind the other to correct astigmatism in ophthalmology~\cite{Elliott2007}.

The organization of the paper is as follows. We first describe the problem of correcting astigmatism and ellipticity in Gaussian beams~(see Sec.~\ref{sec:problem}). We then describe our three-lens solution using a simple analytical model that describes the propagation of Gaussian beams through optical components~(see Sec.~\ref{sec:solution}). 
We show that two counter-rotated cylindrical lenses act as an effective biaxial lens with tunable focal lengths~(see Sec.~\ref{sec:solution_biaxial}) that can correct astigmatism~(see Sec.~\ref{sec:solution_astigmatism}) and ellipticity using a preceding cylindrical lens~(see Sec.~\ref{sec:solution_circularization}). 
We then show that the minimum circularity of the output beam is robust against variations in the separation distance between the two lenses of the pair~(see Sec.~\ref{sec:robustness}). We finally validate our model by mapping the effect of a tunable biaxial lens on a circular beam~(see Sec.~\ref{sec:biaxial}) and by circularizing the output beam of a commercial laser source~(see Sec.~\ref{sec:circularization}).

\section{THEORY -- GAUSSIAN BEAM CIRCULARIZATION}\label{sec:problem}
We seek a simple solution to the problem of converting an astigmatic elliptical beam into a circular Gaussian beam (see Fig.~\ref{fig:gaussian_beams}).
We refer to a circular Gaussian beam (see Fig.~\ref{fig:gaussian_beams}a) as the fundamental TEM00 mode described by the Laguerre-Gaussian polynomials, which are solutions to the Helmholtz equation under the paraxial approximation~\cite{Verdeyen1995}. The electric field vector of a circular Gaussian beam is given by
\begin{equation}
\label{eq:gaussian}
    \bm{E}(x, y, z) = \bm{E}_0 \dfrac{w_0}{w(z)} \exp\left[-\dfrac{x^2 + y^2}{w^2(z)}\right] \exp\left[-jk\dfrac{x^2+y^2}{2R(z)}\right] \exp\left[jkz - j\arctan\left(\dfrac{z}{z_R}\right)\right],
\end{equation}
where $\bm{E}_0$ is the field normalization vector, $w(z)=w_0 \sqrt{1 + (z-z_0)^2/z_R^2}$ is the position-dependent beam radius at position $z$ along the propagation axis, $w_0=w(z_0)$ is the beam radius at the beam waist $z=z_0$, $z_R = \pi w_0^2 / \lambda$ is the Rayleigh range, $\lambda$ is the wavelength, and 
\begin{equation}
    R(z) = (z-z_0) \left[1 + \left(\dfrac{z_R}{z-z_0}\right)^2\right]
\end{equation} 
is the radius of curvature of the phase front. 

The transverse intensity profile of a circular Gaussian beam in any $z$ plane is given by a two-dimensional Gaussian distribution with radial symmetry,
\begin{equation}
    I(x, y, z)=\dfrac{1}{2} \epsilon c |\bm{E}(x, y, z)|^2 = \dfrac{1}{2} \epsilon c |\bm{E}_0|^2 \dfrac{w_0^2}{w^2(z)} \exp\left[-\dfrac{2\left(x^2 + y^2\right)}{w^2(z)}\right],
\end{equation}
where $\epsilon$ is the electric permittivity, and $c$ is the speed of light.

The circular Gaussian beam is an idealization that often fails to describe the output beam of a commercial laser source. For example, as mentioned in the introduction, cavity-based lasers often contain an active medium crystal typically oriented at the Brewster angle to minimize reflection loss. This crystal breaks axial symmetry, resulting in beams whose intensity profiles are not radially symmetric~\cite{Hanna1969, Ramirez2016}.
To describe Gaussian beams with ellipticity and astigmatism~(see Fig.~\ref{fig:gaussian_beams}b-c), we instead rely on a more general solution to the paraxial Helmholtz equation,
\begin{equation}
\begin{split}
\label{eq:generalized_gaussian}
\bm{E}(x, y, z) = \bm{E}_0  \sqrt{\dfrac{w_{0x} w_{0y}}{w_{x}(z) w_{y}(z)}} 
\exp\left[ 
  -\left(\dfrac{x^2}{w_x^2(z)} + \dfrac{y^2}{w_y^2(z)}\right)\right]  
\exp\left[-jk\left(\dfrac{x^2}{2R_x(z)}+\dfrac{y^2}{2R_y(z)}\right)\right]\\
\exp\left[jkz- \dfrac{j}{2}\left(\arctan\left(\dfrac{z - z_{0x}}{z_{Rx}}\right) + \arctan\left(\dfrac{z - z_{0y}}{z_{Ry}}\right)\right)\right],
\end{split}
\end{equation}
where, without loss of generality, the axes of the beam are chosen to be oriented along the $x$ and $y$ axes. Here, $w_{\mu} = w_{0\mu} \sqrt{1 + (z-z_{0\mu})^2/z_{R\mu}^2}$ is the position-dependent beam radius for $\mu \in \{x, y\}$, $z_{R\mu} = \pi w_{0\mu}^2 / \lambda$ is the Rayleigh range, and \begin{equation}
    R_{\mu}(z) =  (z - z_{0\mu})\left[1 + \left(\dfrac{z_{R\mu}}{z - z_{0\mu}}\right)^2\right]    
\end{equation}
is the radius of curvature of the phase front of the beam. The beam waist $z_{0\mu}$ denote the plane in which the beam radius achieves its minimal value, such that $w_\mu(z_{0\mu})=w_{0\mu}$. In the presence of astigmatism, the beam waists $z_{0x}$ and $z_{0y}$ do not overlap (see Fig.~\ref{fig:gaussian_beams}c). 

To further characterize the Gaussian beam, we define its divergence angle, $\alpha_\mu$, as the asymptotic rate of change of the beam radius,
\begin{equation}\label{eq:gaussian_beam_divergence_definition}
    \tan(\alpha_{\mu}) = \lim_{z\rightarrow\infty}\dfrac{d w_{\mu}}{dz} = \dfrac{w_{0\mu}}{z_{R\mu}},\ \mu \in \{x, y\}.
\end{equation}
This equation relies on the fact that, far from the beam waists, the beam radii increase linearly with distance. Equations~\eqref{eq:gaussian} and \eqref{eq:generalized_gaussian} coincide for circular Gaussian beams with equal beam waist radii, $w_{0x} = w_{0y}$, and overlapping beam waists, $z_{0x} = z_{0y}$.

\subsection{Three-Lens Solution to Gaussian Beam Circularization}\label{sec:solution}
Converting an elliptical beam with astigmatism into a circular beam without astigmatism requires suppressing both astigmatism and ellipticity. Suppressing astigmatism requires overlapping the beam waists $z_{0x}$ and $z_{0y}$ (see Fig.~\ref{fig:gaussian_beams}b). In the absence of astigmatism, the resulting beam may still exhibit ellipticity if the beam waist radii differ, $w_{0x} \ne w_{0y}$. Suppressing ellipticity requires equalizing the beam waist radii to achieve $w_{0x} = w_{0y}$. We refer to \emph{beam circularization} as the process of correcting astigmatism and ellipticity. Mathematically, beam circularization involves converting a Gaussian beam described by Eq.~\eqref{eq:generalized_gaussian} into a circular Gaussian beam whose intensity profile is radially symmetric in all planes. 

A typical approach to solving the beam circularization problem is to use two uniaxial cylindrical lenses, $L_1$ and $L_2$ (represented as thick blue and red-orange vertical lines in Fig.~\ref{fig:two_lenses}, respectively). Each lens is aligned with either the major or minor axis of the beam and positioned approximately one focal length away from its corresponding beam waist. The ratio of the focal lengths must be chosen to satisfy 
\begin{equation}\label{eq:two_lens_ratio_approximate_condition}
    f_x / f_y \approx w_{0x}/w_{0y},
\end{equation}
where the exact ratio for Gaussian beams has a complex dependence on $w_{0x}$, $w_{0y}$, $\Delta z = z_{0y} - z_{0x}$, and the wavelength $\lambda$. Achieving this exact ratio is typically challenging in practice when using lenses from standard optics catalogs. These lenses might also deviate from being perfectly uniaxial, introducing unwanted focusing along the weak axis of the lens.

\begin{figure}[t!]
\centering
\includegraphics[]{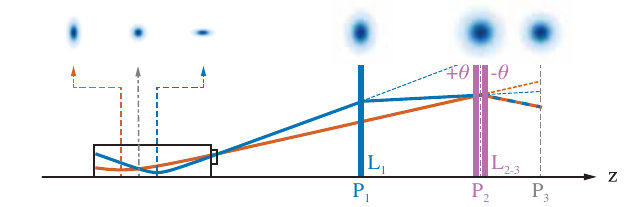}
\caption{
\label{fig:three_lenses}
\textbf{Three-lens solution.}
An astigmatic elliptical Gaussian beam is converted into a circular Gaussian beam using three cylindrical lenses. The first lens ($L_1$) is oriented along the major axis of the beam and placed in an arbitrary plane, not necessarily located a focal length away from the beam waist. The second and third lenses ($L_{2-3}$) are mounted one behind the other at an angle $\pm\theta$ in the plane where the intensity profile of the beam is circular. The relative angle is chosen to equalize the divergence angles of the beam along the major and minor axes. The solution is robust against small changes in the relative distance between the two lenses. The resulting circular beam can readily be collimated using a spherical lens.
}
\end{figure}

Our approach to solving the circularization problem involves three cylindrical lenses~(see Fig.~\ref{fig:three_lenses}). The first cylindrical lens ($L_1$, represented as a thick blue line in Fig.~\ref{fig:three_lenses}) is placed in an arbitrary plane, $P_1$, and oriented along the major axis of the beam (the axis along which the beam divergence is the largest). The lens reduces the divergence of the beam along the major axis so as to create a plane, $P_2$, in which the beam radii along the major and minor axes are equal, i.e., the intensity profile of the beam is circular. The second and third lenses ($L_{2-3}$, represented by thick purples lines in Fig.~\ref{fig:three_lenses}) are mounted one behind the other in the plane $P_2$. Both lenses are aligned along the same axis, then counter-rotated by $\pm\theta$. The relative angle between the two lenses is thus $2\theta$. The relative angle is increased until the beam waists overlap, thereby correcting astigmatism. The resulting beam is a circular Gaussian beam, free of astigmatism and ellipticity. Its intensity profile is radially symmetric in any subsequent $z$ plane. It is readily collimated using standard spherical lenses. 

We now demonstrate the theoretical validity of our approach in three steps. We first show that the cylindrical lens pair acts as an effective biaxial lens with tunable focal lengths~(Sec.~\ref{sec:solution_biaxial}). We then show that this tunable biaxial lens can be used to correct astigmatism~(Sec.~\ref{sec:solution_astigmatism}). We finally show that placing the biaxial lens in a plane where the beam intensity is radially symmetric corrects both ellipticity and astigmatism simultaneously~(Sec.~\ref{sec:solution_circularization}).

\begin{figure}[t!]
\centering
\includegraphics[]{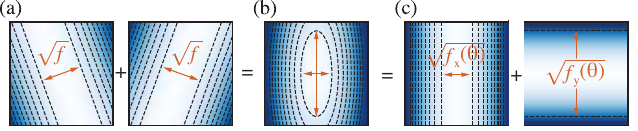}
\caption{
\label{fig:phase_mask}
\textbf{Equivalent phase profiles of a cylindrical lens pair.}
(a)~The cumulative phase profile of two uniaxial cylindrical lenses of focal length $f$, counter-rotated by $\pm\theta$, is equivalent to the phase profile of a (b)~biaxial lens, which is itself equivalent to the cumulative profile of (c)~two uniaxial cylindrical lenses with tunable focal lengths $f_x(\theta)=f\sec^2{(\theta)}/2$ and $f_y(\theta)=f\csc^2{(\theta)}/2$. 
}
\end{figure}

\subsubsection{Tunable Biaxial Lens from Two Uniaxial Cylindrical Lenses}\label{sec:solution_biaxial}

We show that two cylindrical lenses of focal length $f$ counter-rotated by an angle $\pm\theta$ about the $x$ axis act as an effective biaxial lens with tunable focal lengths given by 
\begin{eqnarray}\label{eq:effectve_focal_length_lens_pair}
f_{x}(\theta) &=& f\sec^2(\theta)/2,\\
f_{y}(\theta) &=& f\csc^2(\theta)/2. 
\end{eqnarray}

This result can be obtained by considering the transmittance function of a uniaxial cylindrical lens, which is given by 
\begin{equation}
\mathcal{T}_{x}(f) = \exp\left[ -\dfrac{i\pi}{\lambda f} x ^2 \right]
\end{equation}
when its major axis is oriented along the x axis, and by
\begin{equation}
\mathcal{T}_{\pm\theta}(f) = \exp\left[-\dfrac{i\pi}{\lambda f} \left[ x \cos(\theta) \pm y \sin(\theta) \right]^2 \right]
\end{equation}
when its major axis is rotated by $\pm\theta$ from the $x$ axis. Neglecting the propagation of the beam between the two lenses or, equivalently, assuming a zero inter-lens distance, the cumulative transmittance function of the two cylindrical lenses is given by the product of their individual transmittance functions,
\begin{eqnarray}\label{eq:transmittance_equivalence}
\mathcal{T}_\text{pair}(f,\theta) &=&  \mathcal{T}_{+\theta}(f) ~\mathcal{T}_{-\theta}(f) \\
~ &=& \exp\left[ -\dfrac{i\pi}{\lambda f} \left[ x \cos(\theta) + y \sin(\theta) \right]^2 \right] \exp\left[ -\dfrac{i\pi}{\lambda f} \left[ x \cos(\theta) - y \sin(\theta) \right]^2 \right]
\\
\label{eq:transmittance_equivalence2}
~ &=&
\exp\left[-\dfrac{i\pi}{\lambda (f/2)} x^2 \cos^2(\theta)\right] \exp\left[-\dfrac{i\pi}{\lambda (f/2)} y^2 \sin^2(\theta)\right]\\
\label{eq:transmittance_equivalence3}
~ &=& \mathcal{T}_x \left[\frac{f}{2}{\sec^2(\theta)}\right]~\mathcal{T}_y\left[\frac{f}{2}{\csc^2(\theta)}\right].
\end{eqnarray}
The transmittance function of the cylindrical lens pair is equal to the product of the transmittance functions of two uniaxial lenses oriented along the $x$ and $y$ axes with focal lengths $f_x = f\sec^2(\theta)/2$ and $f_y = f\csc^2(\theta)/2$, respectively. The cylindrical lens pair thus acts as an effective biaxial lens with tunable focal lengths. This result can be understood visually by considering the spatial representation of the phase profiles of the cylindrical lens pair~(see Fig.~\ref{fig:phase_mask}).

\subsubsection{Astigmatism Correction}\label{sec:solution_astigmatism}
\begin{figure}[t!]
\centering
\includegraphics[]{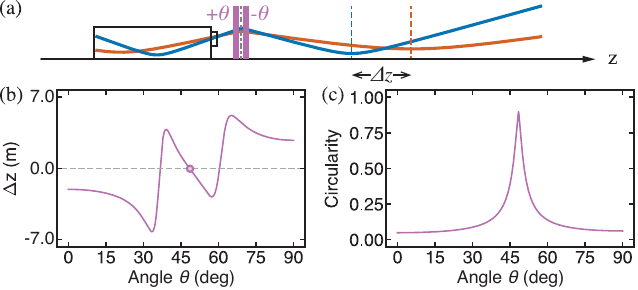}
\caption{
\label{fig:astigmatism}
\textbf{Astigmatism correction.}
(a)~An elliptical beam is focused through a pair of cylindrical lenses with a tunable relative angle $2\theta$. The major and minor axes focus in different planes (blue and orange dashed lines).
(b)~The relative distance between the two focal planes (see Eq.~\eqref{eq:relative_distance_beam_waists}) changes with the lens angle and vanishes at three specific values.
(c)~Beam circularity (see Eq.~\eqref{eq:fom}) is maximized close to one of the angles where the focal planes overlap; however, the biaxial lens pair alone is not sufficient to achieve perfect circularity. Note that the angle for which the circularity is maximized does not necessarily coincide with the angle for which astigmatism is suppressed, and the plots are not necessarily symmetric.
}
\end{figure}

The biaxial lens can suppress astigmatism, converting an elliptical beam with astigmatism into an elliptical beam without astigmatism (see Fig.~\ref{fig:astigmatism}). By rotating the relative angle between the two cylindrical lenses, the focal lengths of the effective biaxial lens is chosen to minimize the relative separation between the beam waists, denoted $\Delta z$ (see Fig.~\ref{fig:astigmatism}b). The relative separation between the beam waists is defined as
\begin{equation}\label{eq:relative_distance_beam_waists}
    \Delta z(\theta) = z_{0y}(\theta) - z_{0x}(\theta),   
\end{equation}
where the dependence of the beam waist on $\theta$ can be calculated analytically (see Eq.~\eqref{eq:gaussian_after_lens_z0} in App.~\ref{sec:gaussian_lens_properties_derivation}).
For sufficiently small values of the focal length $|f|$, there exists at least one angle $\theta^*$ for which the two beam waists coincide, i.e., $\Delta z(\theta^*) = 0$ (see App.~\ref{sec:derivation_doublet_deastigmatization} for a proof of existence of $\theta^*$). This statement means that the lens pair can always be used to correct astigmatism in any input Gaussian beam.

Even after overlapping the beam waists, the waist radii may still differ, and some ellipticity may remain (see Fig.~\ref{fig:astigmatism}c). This limitation arises because a thin lens can only modulate the phase of the input beam without directly modifying its amplitude profile. Although phase modulation can influence the intensity profile of the beam upon propagation, it cannot, in general, eliminate radial asymmetry in the immediate vicinity of the lenses. As a result, the output beam remains an elliptical Gaussian beam, unless the beam has a circular intensity profile at the input of the lens pair. Making the intensity profile of the input beam circular requires the use of another cylindrical lens before the lens pair.

\subsubsection{Ellipticity Correction}\label{sec:solution_circularization}

Suppressing ellipticity requires placing the cylindrical lens pair in a plane $P_2$ where the intensity profile of the beam is circular. This plane can be found by placing a uniaxial cylindrical lens oriented along the major axis of the beam in a plane $P_1$ preceding $P_2$. This lens reduces the divergence of the beam along the major axis of the beam such that, for sufficiently small focal lengths $f_1>0$, there exists a plane $P_2$ in which the beam radii along the minor and the major axes overlap (see App.~\ref{sec:first_lens_existence} for a proof of existence for $f_1$).

Given that the focal lengths of the cylindrical lenses of the lens pair are equal, i.e., $f_2 = f_3 = f$, the optimal rotation angle at which the beam is circular is given by: 
\begin{equation}
    \label{eq:circularization_angle}
    \theta^* = \dfrac{1}{2} \text{arccos}\left[ \dfrac{f}{2} \left(\dfrac{\lambda}{\pi w_r}\right)^2 \left\{ \dfrac{z_{0y}}{w_{0y}^2} - \dfrac{z_{0x}}{w_{0x}^2} \right\} \right],
\end{equation}
where $w_{0\mu}$ is the beam waist radius of the input beam for $\mu\in\{x,y\}$, $z_{0\mu}$ is the beam waist, and $w_r = w_{x}(z_{P_2}) = w_{y}(z_{P_2})$ is the beam radius at $P_2$ (see App.~\ref{sec:derivation_doublet_circularization} for a full derivation). This equation is valid as long as
\begin{equation}
    \label{eq:circularization_focal_length_condition}
    |f| \leq f_{\max} = \bigg\rvert \dfrac{2 \left(\frac{\pi w_r}{\lambda}\right)^2 }{\frac{z_{0y}}{w_{0y}^2} - \frac{z_{0x}}{w_{0x}^2}} \bigg\rvert.
\end{equation}
When $f$ satisfies this inequality, an astigmatic elliptical beam is converted into a circular beam with overlapping beam waists, $z_{0x}(\theta^*) = z_{0y}(\theta^*)$, and beam waist radii, $w_{0x}(\theta^*) = w_{0y}(\theta^*)$. When $f$ violates this inequality, no real-valued solution for $\theta^*$ exists, and circularization cannot be achieved. 

The above results are valid when the two lenses of the cylindrical lens pair are perfectly uniaxial. In practice, these lenses may have a finite focal length $f_{\perp}$ along their weak axis due to manufacturing imperfections. We can prove that for a finite $f_{\perp}$, Eqs.~\eqref{eq:circularization_angle} and \eqref{eq:circularization_focal_length_condition} are still valid when replacing the nominal focal length $f$ by the effective focal length (see App.~\ref{sec:circularization_non_uniaxial_lens})
\begin{equation}
    f' = \left(\dfrac{1}{f} - \dfrac{1}{f_{\perp}}\right)^{-1}.
\end{equation}
This result is valid when both lenses have the same transverse focal lengths, $f_\perp=f_{2\perp} = f_{3\perp}$. However, when the two focal lengths are not equal, $f_{2\perp} \neq f_{3\perp}$ the beam at the output of the lens pair exhibits residual ellipticity and astigmatism for all angles $\theta$, though these effects are suppressed compared to the typical two-cylindrical-lens solution (see App.~\ref{sec:circularization_non_uniaxial_lens} for details).

\begin{figure}[t!]
\centering
\includegraphics[]{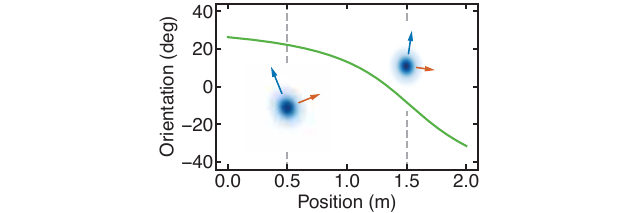}
\caption{
\label{fig:generalized_gaussian_twisting}
\textbf{Generalized Gaussian beam twisting.}
The orientation of a generalized Gaussian beam rotates as it propagates through free space, while its transverse intensity profile in each plane is elliptical, similar to the intensity profile of an elliptical Gaussian beam (see insets).
}
\end{figure}

\subsection{Robustness Against Finite Inter-Lens Distance}\label{sec:robustness}
Our calculations have so far assumed a negligible distance between the two lenses of the cylindrical lens pair. In practice, this distance is lower-bounded by the thickness of the lenses and the rotation mounts to which they are attached. 
A nonzero distance between the two cylindrical lenses results in a breakdown of the transmittance results presented in Sec.~\ref{sec:solution_biaxial}. In this case, the lens pair no longer acts as an effective biaxial lens, and the beam after the lens pair can no longer be accurately modeled by a standard Gaussian beam~(see Eq.~\eqref{eq:generalized_gaussian}).

To study the effects of the nonzero distance between the two cylindrical lenses, a generalized model for beam propagation is required. We now introduce this model and use it to quantify the degradation in beam circularity resulting from variations in the inter-lens distance. We show that the imperfections introduced by the nonzero inter-lens distance can be mitigated by using a large focal length $f$ for the two cylindrical lenses.

\subsubsection{Generalized Gaussian Beams}
A sufficient model to describe Gaussian beams passing through cylindrical lenses with arbitrary orientations is given by 
\begin{equation}\label{eq:generalized_gaussian_description}
    \bm{E}(x, y, z) = \bm{E_0}(z) \exp\left(-\bm{r}^T \bm{\Lambda}(z) \bm{r}\right),
\end{equation}
where $\bm{E_0}(z) \in \mathbb{C}^{2}$ is the complex-valued electric field amplitude, $\bm{r} = \left(x, y\right)^T$ is the transverse coordinate vector, and $\bm{\Lambda}(z) \in \mathbb{C}^{2 \times 2}$ is a complex-valued $2 \times 2$ matrix (see App.~\ref{sec:matrix_formalism} for a detailed presentation). We refer to beams described by Eq.~\eqref{eq:generalized_gaussian_description} as \emph{generalized Gaussian beams}.

Gaussian beams with ellipticity and astigmatism are a special case of generalized Gaussian beams for which the commutator of $\Re\bm{\Lambda}(z)$ and $\Im\bm{\Lambda}(z)$ is zero, i.e., $\left[\Re\bm{\Lambda}(z), \Im\bm{\Lambda}(z)\right] = 0$. When Gaussian beams pass through a cylindrical lens oriented along an arbitrary axis, they become non-Gaussian beams with $[\Re\bm{\Lambda}(z), \Im\bm{\Lambda}(z)] \neq 0$. Despite maintaining an intensity profile identical to that of a Gaussian beam, the nonzero commutator gives rise to a ``twisting'' effect, wherein the principal axes of the beam rotate during free-space propagation~(see Fig.~\ref{fig:generalized_gaussian_twisting}).

\begin{figure}[t!]
\centering
\includegraphics[]{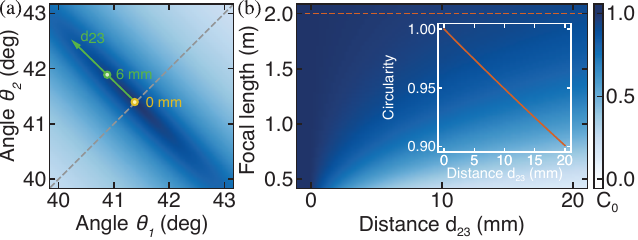}
\caption{
\label{fig:imperfections}
\textbf{Dependence of minimum circularity on nonzero inter-lens spacing.}
(a)~The minimum circularity $\mathcal{C}_0$ (white-to-blue surface) changes with the rotation angle of the cylindrical lenses, $\theta_2$ and $\theta_3$. At $d_{23}=0$ (yellow circle), the optimal minimum circularity is achieved for $\theta_2=\theta_3=\theta^*$ (gray dashed line), where $\theta^*=41.4^\circ$. For $d_{23}\neq0$, the optimal angles deviates from the optimal predicted values (green line). The green disk indicates the optimal angles obtained for $d_{23} = 6\ \text{mm}$, which approximately corresponds to the distance between the lenses when mounted in rotation mounts.
(b)~The minimum circularity of the beam depends on both the focal length of the lenses and their inter-lens distance. The circularity decreases with inter-lens spacing. For a fixed inter-lens spacing, increasing the focal length improves the minimum achievable circularity. These results were obtained for a Gaussian beam with a wavelength of $\lambda = 415~\text{nm}$ whose beam waist radii are $w_{0x}=260~\mu\text{m}$ and $w_{0y}=1.015~\text{mm}$ at the beam waists located $z_{0x}=-6~\text{m}$ and $z_{0y}=-22~\text{m}$, respectively, where $z=0$ denotes the plane where the first lens of the lens pair is located.
}
\end{figure}

\subsubsection{Dependence of Beam Circularity on Inter-Lens Distance}

The generalized Gaussian beam model provides a means to study changes in the intensity profile of the beam as a function of the inter-lens distance. By computing the evolution of the matrix $\bm{\Lambda}(z)$ with respect to $z$ (see Table~\ref{table:lambda_rules}), we can determine the beam radii and orientation in all planes $z$ by diagonalizing $\Re\bm{\Lambda}(z)$: the principal axes of the beam are oriented along the eigenvectors of $\Re\bm{\Lambda}(z)$, whereas the beam radii are given by $w_j = 1/\sqrt{\lambda_j}$, where $\lambda_j$ are the associated eigenvalues for $j = 1, 2$. We note that the eigenvectors are not necessarily aligned with the lab-frame axes, $x$ and $y$, and may vary with propagation distance due to beam twisting.

We define the \emph{circularity} of a generalized Gaussian beam in the plane $z$ as the ratio of the smaller radius to the larger radius of the beam in that plane, 
\begin{equation}
\mathcal{C}(z)=\frac{\min(w_1(z), w_2(z))}{\max(w_1(z), w_2(z))}. 
\end{equation}
We then define the \emph{minimum circularity} of a beam as
\begin{equation}
    \label{eq:fom}
    \mathcal{C}_0 = \inf_{z \geq z_3} \mathcal{C}(z).
\end{equation}
This quantity represents the infimum of $\mathcal{C}(z)$ over all $z$ beyond the position of the third lens of the three-lens system, located at $z_3 = z_{P_2}+d_{23}/2$, assuming the beam propagates indefinitely in free space.

We numerically compute the minimum circularity of a generalized Gaussian beam for different values of inter-lens distance, $d_{23}$, as a function of the rotation angles $\theta_2$ and $\theta_3$ of the two cylindrical lenses with respect to the $x$ axis (see Fig.~\ref{fig:imperfections}a). 
The input beam is chosen to be an elliptical Gaussian beam with astigmatism that is circular in the plane $P_2$ where the lens pair is placed. The focal length of the two cylindrical lenses is chosen to be $f=2000~\text{mm}$.

When the inter-lens distance is zero (see yellow disk in Fig.~\ref{fig:imperfections}a for $d_{23}=0$), the optimal minimum circularity is obtained for a specific value of $\theta^*=41.4^\circ$ located along the main diagonal defined by $\theta = \theta_2 = \theta_3$ (see gray dashed line). Deviation away from this optimal angle leads to degradation in the minimum circularity (blue-to-white shaded surface). As the inter-lens distance increases (see green line in Fig.~\ref{fig:imperfections}a with the green marker corresponding to $d_{23}=6~\text{mm}$), the optimal angles for reaching the optimal minimum circularity are no longer equal $\theta_2^*\neq\theta_3^*$. 

We further compute the change in minimum circularity for different focal lengths and inter-lens distances, $d_{23}$ (see Fig.~\ref{fig:imperfections}b). The minimum circularity is computed for the case where $\theta_2=\theta_3$. For $d_{23} = 0$, the circularity of the beam is equalt for all focal lengths. As $d_{23}$ departs from $0$, the minimum circularity decreases approximately linearly (see inset in Fig.~\ref{fig:imperfections}b). The loss of circularity with increasing inter-lens distance is more pronounced for smaller focal lengths.
We conclude that, for the specific test beam used in our simulations, mitigating the detrimental effects of inter-lens distance requires choosing a large focal length $f$, ideally approaching its upper bound $f_\text{max}$ defined in the right hand side of Eq.~\eqref{eq:circularization_focal_length_condition}. This conclusion is generally valid for any input beam with a sufficiently small divergence angle. Indeed, larger focal lengths correspond to weaker lenses, which cause a slower evolution of the beam’s transverse profile across the inter-lens distance. As a result, the beam properties remain approximately unchanged between the lenses. In contrast, smaller focal lengths induce rapid variations in the beam radii between the two lenses, increasing the sensitivity to inter-lens spacing and ultimately degrading the circularization performance. However, the optimal choice of $f$ should generally be checked for any specific beam, especially for beams with large divergence angles.


\begin{figure}[t!]
\centering
\includegraphics[]{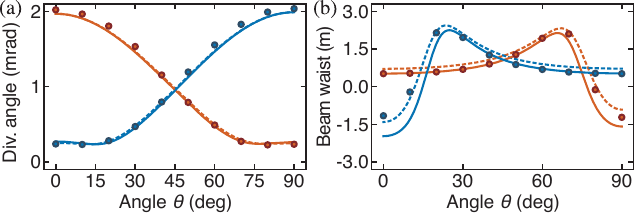}
\caption{
\label{fig:validation}
\textbf{Model validation.}
(a)~Divergence angles and (b)~beam waists of an astigmatic elliptical Gaussian beam measured along its principal axes, which are aligned with the $x$ (red) and $y$ axes (blue). The beam is generated by transmitting a nearly circular Gaussian beam at $780~\text{nm}$ through a pair of cylindrical lenses with a tunable relative angle $2\theta$, oriented asymmetrically about the $x$ axis. Each data point represents a fit of the longitudinal intensity profile to a Gaussian beam model. Solid lines indicate theoretical predictions based on the input beam and the geometry of the lens system. The cylindrical lenses have a focal length of $f = 1000~\text{mm}$. The dashed lines show the expected behavior obtained when choosing the lenses to have an effective focal length along the transverse direction $f_{\perp} = 51.4~\text{m}$.
}
\end{figure}

\section{EXPERIMENT -- ASTIGMATISM AND ELLIPTICITY SUPPRESSION}\label{sec:experiment}

Having established the validity and robustness of our method, we now demonstrate its applicability in an experimental setting. We begin by confirming that a pair of cylindrical lenses acts as a biaxial lens with tunable focal lengths. We then apply our method to circularize the beam emitted by a commercial titanium:sapphire laser source.

\subsection{Biaxial Lens with Tunable Focal Lengths}\label{sec:biaxial}
To experimentally confirm that a pair of counter-rotated cylindrical lenses acts as an effective biaxial lens with tunable focal lengths, we measured the changes in the properties of a circular Gaussian beam and compared them against those computed using the transmittance function of Eqs.~\eqref{eq:transmittance_equivalence}--\eqref{eq:transmittance_equivalence3}. 

We created a collimated circular Gaussian beam using a fiber collimator at $780~\text{nm}$. The beam had a minimum circularity of $\mathcal{C}_0 \geq 0.95$ with a collimated beam radius of $1.1~\text{mm}$ and a beam divergence of $0.1~\text{mrad}$. We then placed the cylindrical lens pair at an arbitrary distance of $240~\text{mm}$ away from the fiber collimator. The two cylindrical lenses, each with a focal length of $f = 1000~\text{mm}$ (Thorlabs, LJ4530RM), were mounted in a rotation mount (Thorlabs, CRM1T) and oriented parallel to the optical table along the $x$ axis. We counter-rotated each lens by an angle $\pm\theta$ and measured the intensity profile of the focusing beam at various longitudinal planes near its beam waists (see App.~\ref{sec:image_analysis} for details on the image acquisition and processing procedure). We then fit each transverse intensity profile to an elliptical Gaussian beam with beam radii
\begin{equation}
    w_{\mu}(z) = w_{0\mu}' \sqrt{1 + \left((z-z'_{0\mu}) \dfrac{\lambda}{\pi w_{0\mu}'^2}\right)^2},~\mu\in\{x,y\}
\end{equation}
to obtain the beam waists $z_{0\mu}$ along each axis, as well as the beam divergence, $\alpha$ (see Eq.~\eqref{eq:gaussian_beam_divergence_definition}).


The measurements of the divergence angles and beam waists are in good agreement with theoretical predictions (see solid lines on Fig.~\ref{fig:validation}), except for small deviations in the beam waist locations near $\theta = 0^\circ$ and $\theta = 90^\circ$. At $\theta = 0^\circ$, the lens pair should ideally have no effect on the beam along the $y$ axis, as both lenses are nominally aligned along the $x$ axis. The observed deviation from theory suggests a shift in the beam waist along the $y$ axis at $\theta = 0^\circ$. This discrepancy cannot be attributed to inaccurate lens orientation alone. Our calculations show that an angular misalignment of up to $5^{\circ}$ would be required to produce such an effect, which is substantially greater than the alignment precision of less than $1^{\circ}$ for the our rotation mounts. We attribute the observed discrepancy to the cylindrical lenses not being perfectly uniaxial, exhibiting a small residual focusing power along their transverse axis. The discrepancy between theory and experiment is minimized when we assume that the two lenses of the lens pair have a residual transverse focal length of $f_{\perp} = 51.4~\text{m}$ along their weak axis (see dashed lines in Fig.~\ref{fig:validation}). A similar explanation accounts for the deviation observed near $\theta = 90^\circ$. These experimental data confirm that the lens pair acts as an effective biaxial lens system with tunable focal lengths.

\begin{figure}[t!]
\centering
\includegraphics[]{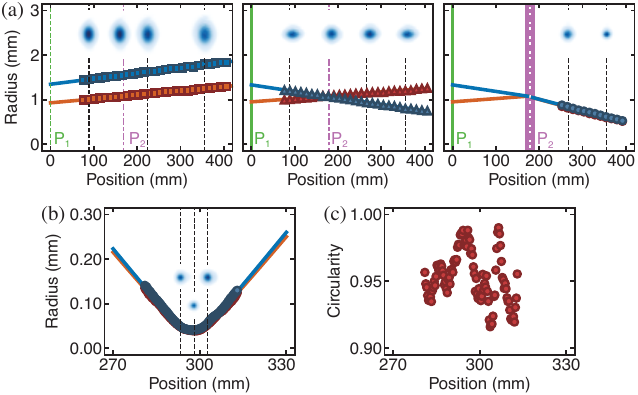}
\caption{
\label{fig:exp_2}
\textbf{Experimental demonstration of beam circularization.}
(a-left)~A commercial laser source emits a diverging beam with an elliptical intensity profiles (insets). The beam radii (squares) are measured along the $x$ (red) and $y$ (blue) axes and fitted to Gaussian beam profiles (solid lines).
(a-middle)~Focusing the beam along its major axis using a cylindrical lens ($f_1=500~\text{mm}$) placed at $P_1$ (thick green line) creates a plane $P_2$ at $z = 176~\text{mm}$ where the intensity profile becomes circular.
(a-right)~Focusing the beam along both its major and minor axes using a cylindrical lens pair ($f_2 = f_3 = 500~\text{mm}$) with an optimized relative angle suppresses astigmatism and restores circularity.
(b)~Focusing the beam after $P_2$ with an achromatic doublet ($f = 150~\text{mm}$) further demonstrates the suppression of astigmatism, with the beam waists overlapping at $z = 298~\text{mm}$.
(c)~The minimum beam circularity near the beam waist averages to $0.95(1)$, bounded between $0.92(1)$ and $0.99(1)$.
}
\end{figure}

\subsection{Beam Circularization of a Commercial Laser Source}\label{sec:circularization}

We now show that the three-lens system can be used to circularize the output of a commercial titanium:sapphire laser source at $813~\text{nm}$.
The laser source emits a diverging beam with an elliptical intensity profile whose principal axes align with the $x$ and $y$ directions of the lab frame, defined as parallel and perpendicular to the optical table.  The circularity of the beam in the far-field is $\mathcal{C}_\infty=0.69$ (see Fig.~\ref{fig:exp_2}a). 


We placed the first cylindrical lens with focal length $f_1=500~\text{mm}$ (Thorlabs, LJ4147RM) at a distance of $1300~\text{mm}$ away from the output window of the laser. 
This location was chosen arbitrarily to satisfy the constraints of our scientific apparatus while providing sufficient space for beam characterization. 
The lens was mounted in a rotation mount (Thorlabs, CRM1PT) and oriented along the major axis of the elliptical beam. Because of the properties of generalized Gaussian beams, whose principal axes might undergo an anti-crossing effect (see App.~\ref{sec:imperfect_orientation_anti_crossing}),any relative orientation misalignment between the lens and the major axis results in the beam circularity to be strictly less than $1$ in all subsequent planes after the lens. To minimize misalignment errors, we measured the beam profile across a range surrounding the expected location of $P_2$ and adjusted the orientation to maximize the circularity within this range.
After optimization, we measured a circularity of  $\mathcal{C}_{P_2}=0.9996^{+0.0004}_{-0.0050}$ in the plane $P_2$ located $z = 186~\text{mm}$ away from the lens~(see Fig.~\ref{fig:exp_2}b).

We then placed two plano-convex cylindrical lenses with equal focal length $f = 500~\text{mm}$ (Thorlabs, LJ4147RM) symmetrically around the plane $P_2$, each offset by $\pm3~\text{mm}$ along the optical axis ($d_{23} = 6~\text{mm}$). The lenses were mounted in rotation mounts (Thorlabs, CRM1PT) with their convex surfaces facing the incoming beam. Their major axes were aligned with the beam's minor $x$-axis and counter-rotated by an angle $\pm\theta$ to correct astigmatism. After optimization, the circularity of the beam in the far field was measured to be $\mathcal{C}_{\infty} = 0.97 \pm 0.01$ at the optimal lens angle of $\theta^* = 28.7^{\circ}$. We confirmed that the beam maintained a circular intensity profile throughout the entire propagation range beyond $P_2$ (see Fig.~\ref{fig:exp_2}c).

To confirm that the lens pair corrected not only ellipticity far from the beam waists but also astigmatism, we focused the beam using an achromatic spherical lens with focal length $f = 150~\text{mm}$ (Thorlabs, AC254-150) at a distance of $30~\text{mm}$ away from $P_2$. We measured the intensity profile near the beam waist to extract the relative distance between the beam waists, $\Delta z$. We observed that this relative distance is highly sensitive to the orientation angle of the cylindrical lenses, with angular deviations as small as $0.05^{\circ}$ resulting in measurable changes. After optimization, we achieved $|\Delta z|=0.008 z_R$, less than $0.8\%$ of the Rayleigh range.

These results demonstrate that our three-lens solution can be used to circularize the elliptical laser beam of a commercial laser source, transforming a highly elliptical Gaussian beam into a nearly circular one. We attribute the residual imperfections to imperfect lens orientation, nonzero inter-lens spacing, and other minor unaccounted factors.

\section{CONCLUSION}
In conclusion, we have introduced a simple three-lens solution to convert astigmatic, elliptical Gaussian beams into circular Gaussian beams without astigmatism. We have validated our solution using theoretical results, numerical simulations, and experimental demonstrations. We have shown that its optimality and robustness to imperfections can be understood through the theory of generalized Gaussian beams, which explains phenomena such as beam twisting and axis anti-crossings.
Our three-lens solution can be readily used to suppress astigmatism and ellipticity in laser beams emitted from commercial laser sources, as well as those degraded by dispersive or imperfect optical elements. Its key advantage over the typical two-lens solution is that it does not impose strict constraints on the choice of lenses. It readily supports many applications that require tunable beam shaping, such as creating large arrays of optical traps with spatially homogeneous properties~\cite{Schymik2022, Chew2024}, minimizing cross-talk in site-selective quantum gate operations~\cite{BinaiMotlagh2023, Lim2025}, and coupling laser beams into optical fibers with high efficiency~\cite{Wagner1982, Sanchez2016}. 
More broadly, any spherical lens in an optical system can be replaced with a tunable cylindrical lens pair to correct residual ellipticity and astigmatism.



\section{ACKNOWLEDGEMENT}
This research was funded thanks in part to CFREF. We acknowledge the support of the Natural Sciences and Engineering Research Council of Canada (NSERC).

\smallskip
\textbf{Disclosures.} The authors declare no competing interests.
\smallskip

\textbf{Data availability.} Data underlying the results presented in this paper are not publicly available at this time but may be obtained from the authors upon reasonable request.


\begin{appendix}

\section{MATRIX FORMALISM FOR GENERALIZED GAUSSIAN BEAMS}\label{sec:matrix_formalism}
Our beam circularization method uses cylindrical lenses with arbitrary orientations to shape the spatial profile of Gaussian beams. After passing through a cylindrical lens with an arbitrary orientation, the beam is no longer Gaussian and cannot be described by Eq.~\eqref{eq:generalized_gaussian}. To describe the propagation of beams through optical systems composed of cylindrical lenses with arbitrary orientations, we introduce generalized Gaussian beams, also called Gaussian beams with general astigmatism~\cite{Arnaud1969}.

Using a matrix representation, we show that the computation of the beam shape reduces to the diagonalization of a $2 \times 2$ complex-valued and symmetric matrix $\bm{\Lambda}(z)$. This matrix depends on the position $z$ along the propagation axis of the beam. We derive a set of rules that determine how $\bm{\Lambda}(z)$ changes with $z$ as the beam propagates in free-space and transmits as it passes through cylindrical lenses with arbitrary orientation. We use this matrix formalism to analyze the shape of the beam through various optical systems.


\subsection{Generalized Gaussian beams}\label{sec:matrix_formalism_intro}

A generalized Gaussian beam can be thought of as a Gaussian beam with an orientation angle $\phi$ that is complex-valued~\cite{Arnaud1969}. Its complex electric field is given by 
\begin{equation}
\label{eq:arnaud_generalized_gaussian}
\begin{split}
    \bm{E}(x, y, z) = \bm{E}_0  \sqrt{\dfrac{w_{0x} w_{0y}}{w_{x}(z) w_{y}(z)}} 
\exp\left[ 
  -\left(\dfrac{\mu(\phi)^2}{w_x^2(z)} + \dfrac{\nu(\phi)^2}{w_y^2(z)}\right)\right]  
\exp\left[-jk\left(\dfrac{\mu(\phi)^2}{2R_x(z)}+\dfrac{\nu(\phi)^2}{2R_y(z)}\right)\right]\\
\exp\left[jkz- \dfrac{j}{2}\left(\arctan\left(\dfrac{z - z_{0x}}{z_{Rx}}\right) + \arctan\left(\dfrac{z - z_{0y}}{z_{Ry}}\right)\right)\right],
\end{split}
\end{equation}
where $\mu(\phi) = x \cos(\phi) - y \sin(\phi)$ and $\nu(\phi) = x \sin(\phi) + y \cos(\phi)$. 
This equation is a valid solution to Maxwell's equations even for complex values of $\phi$. It accurately describes the beam transmitted through an arbitrarily oriented cylindrical lens, where the orientation of the beam can transform from a real-valued angle before the lens to a complex-valued angle afterwards. The complex-valued orientation results in non-Gaussian features in the beam. For instance, while the intensity profile takes the form of a radially asymmetric Gaussian distribution in all planes, the principal axes of the beam can rotate as the beam propagates in free space (see Fig.~\ref{fig:generalized_gaussian_twisting}). We refer to this phenomenon as \textit{twisting}.

The complex electric field given in Eq.~\eqref{eq:arnaud_generalized_gaussian} can be rewritten in the following compact form:
\begin{equation}\label{eq:gaussian_generalized}
\bm{E}(\bm{r}, z) = \bm{A}(z) \exp{\left(-\bm{r}^T \bm{\Lambda}(z) \bm{r}\right)},
\end{equation}
where $\bm{r} = (x, y)^T$ is the transverse position vector, $\bm{\Lambda}(z) \in \mathbb{C}^{2 \times 2}$ is a symmetric rank-2 tensor represented as a two-dimensional matrix with complex coefficients, and $\bm{A}(z)$ is a position-dependent complex-valued envelope function independent of the transverse coordinates $x$ and $y$. 
The associated intensity profile of the generalized Gaussian beam in each plane $z$ is given by
\begin{equation}
\label{eq:generalized_gaussian_intensity}
    I(\bm{r}, z) = \dfrac{1}{2} \epsilon c |\bm{A}(z)|^2 \exp\left(-2 \bm{r}^T~\Re\bm{\Lambda}(z)~\bm{r}\right).
\end{equation}

The $\bm{\Lambda}(z)$ matrix contains sufficient information to determine the beam radii and orientation of the generalized Gaussian beam in all planes $z$. The principal axes of the beam, $\bm{v}_1(z)$ and $\bm{v}_2(z)$, are given by the eigenvectors of $\Re{\bm{\Lambda}(z)}$, and the beam radii are given by $w_\mu(z) = 1/\sqrt{\lambda_{\mu}(z)}$, where $\lambda_{\mu}(z)$ are the eigenvalues of $\Re{\bm{\Lambda}(z)}$, for $\mu \in {1, 2}$.

\subsection{Matrix transformations for computing beam propagation through optical components}\label{sec:lambda_table}

We now show that the computation of $\bm{\Lambda}(z)$ can be performed independently of $\bm{A}(z)$ using a simple set of rules involving only inversion and addition of $2 \times 2$ matrices (see App.~\ref{sec:lambda_table}). This set of rules describes how $\bm{\Lambda}(z)$ evolves under free-space propagation and transmission through cylindrical lenses, assuming the thin-lens approximation (see Table~\ref{table:lambda_rules} for a summary of these rules). Each rule is derived using a matrix formalism analogous to the ABCD matrix method~\cite{Siegman1986}. This formalism eliminates the need to compute the convolution of the electric field of the beam with the free-space propagation kernel from Fresnel diffraction theory~\cite{Goodman2005}, thereby reducing both the computational cost and the analytical complexity of computing the propagation of generalized Gaussian beams through optical systems comprising multiple cylindrical lenses.

\renewcommand{\arraystretch}{1.4} 

\begin{table}[t!]
\centering
\begin{tabularx}{4.25in}{|>{\raggedright\arraybackslash}m{3.5cm}|>{\raggedright\arraybackslash}X|}
\hline
\textbf{Evolution} & \textbf{Rule} \\ \hline

Free-space propagation by $\Delta z$. & 
{\small
$
\begin{aligned}
\bm{\Lambda}(z + \Delta z) = {} & \left(\bm{\Lambda}(z)^{-1} \right.
\left. + \dfrac{i \lambda \Delta z}{\pi} \mathbb{I}_{2 \times 2}\right)^{-1}
\end{aligned}
$
} \\ \hline

Transmission through a spherical lens with focal length $f$. & 
{\small
$
\bm{\Lambda}(z^+) = \bm{\Lambda}(z^-) + \dfrac{i \pi}{\lambda f} \mathbb{I}_{2 \times 2}
$
} \\ \hline

Transmission through a cylindrical lens with focal length $f$ rotate, with the major axis rotated at an angle $\theta$ from the $x$ axis. & 
{\small
$
\begin{aligned}
\bm{\Lambda}(z^+) = \bm{\Lambda}(z^-) + \dfrac{i \pi}{\lambda f} 
\begin{bmatrix}
\cos^2(\theta) & -\frac{1}{2} \sin(2\theta) \\
-\frac{1}{2} \sin(2\theta) & \sin^2(\theta)
\end{bmatrix}
\end{aligned}
$
} \\ \hline

Transmission through a biaxial lens with major focal length $f_1$ and minor focal length $f_2$, with the major axis rotated at an angle $\theta$ from the $x$ axis. & 
{\small
$
\begin{aligned}
\bm{\Lambda}(z^+) = \bm{\Lambda}(z^-) &+ \dfrac{i \pi}{\lambda f_1} 
\begin{bmatrix}
\cos^2(\theta) & -\frac{1}{2} \sin(2\theta) \\
-\frac{1}{2} \sin(2\theta) & \sin^2(\theta)
\end{bmatrix} \\
&+ \dfrac{i \pi}{\lambda f_2} 
\begin{bmatrix}
\sin^2(\theta) & +\frac{1}{2} \sin(2\theta) \\
+\frac{1}{2} \sin(2\theta) & \cos^2(\theta)
\end{bmatrix}
\end{aligned}
$
} \\ \hline

\end{tabularx}
\caption{
\textbf{Transformation rules for generalized Gaussian beams.} Rules governing the evolution of the $\bm{\Lambda}(z)$ matrix under free space propagation and arbitrarily oriented biaxial lenses. $\bm{\Lambda}(z)$ completely describes the intensity profile of generalized Gaussian beams, making these rules sufficient for determining the beam shape of a generalized Gaussian beam.
}
\label{table:lambda_rules}
\end{table}


\subsubsection{Propagation of Gaussian Beams in Free Space}
The free-space propagation of a Gaussian beam can be computed using Fresnel diffraction theory~\cite{Goodman2005}. Given $\bm{E}(x, y, z=0)$, the complex electric field at $z=\Delta z$ is given by
\begin{equation}
    \label{eq:fresnel_diffraction}
    \bm{E}(x, y, z=\Delta z) = \mathcal{F}^{-1}\left[ \mathcal{F}\left(\bm{E}(x, y, z=0)\right) \cdot H_{\Delta z}(x,y) \right],
\end{equation}
where 
\begin{equation}
    \label{eq:freespace_evolution_fourier_transform}
    H_{\Delta z}(f_x, f_y) =  e^{ikz} \exp\left(-i \pi \lambda \Delta z    (f_x^2 + f_y^2) \right)
\end{equation}
is the Fourier transform of the free space evolution kernel~\cite{Goodman2005}, and 
\begin{equation}
    \mathcal{F}\left[\bm{E}(x, y, z0)\right](f_x, f_y) = \iint \bm{E}(x, y, z=0)\exp\left(-2 \pi if_x x - 2 \pi i f_y y\right)~dx~dy,
\end{equation}
is the Fourier transform with respect to the $x$ and $y$ degrees of freedom. 

Given the Fourier transform of a generalized Gaussian beam,
\begin{equation}
    \label{eq:fourier_transform_generalized_gaussian_beam}
    \mathcal{F}\left[E(\bm{r}, z=0) \right](f_x, f_y) = \dfrac{\pi}{\sqrt{\text{det}(\bm{\Lambda}(0))}}\bm{A}(0) \exp\left(-\pi^2 \bm{f}^T \bm{\Lambda}^{-1}(0) \bm{f} \right),
\end{equation}
where $\bm{f} = (f_x, f_y)^T$, we can compute the free-space propagation of a generalized Gaussian beam,
\begin{equation}
    \label{eq:generalized_gaussian_free_space}
    E(\bm{r}, z=\Delta z) =  e^{ik \Delta z} \bm{A}(0)  \cdot \sqrt{\dfrac{\text{det}(\bm{\Lambda}(\Delta z))}{\text{det}(\bm{\Lambda}(0))}} \exp\left(-\bm{r}^T \bm{\Lambda}(\Delta z) \bm{r}\right), 
\end{equation}
where 
\begin{equation}
    \label{eq:lambda_matrix_free_space}
    \bm{\Lambda}(\Delta z) = \left(\bm{\Lambda}(0)^{-1} + \dfrac{i\lambda \Delta z}{\pi} \mathbb{I}_{2 \times 2} \right)^{-1},
\end{equation}
and $\mathbb{I}_{2 \times 2}$ is the 2-by-2 identity matrix. In the case of a Gaussian beam, the coefficient $\sqrt{\det(\bm{\Lambda}(\Delta z)) / \det(\bm{\Lambda}(0))}$ accounts for both the Gouy phase shift and a corrective factor to ensure power conservation. The intensity profile and phase curvature of the beam are fully described by the $\exp(-\bm{r}^T \bm{\Lambda}(\Delta z) \bm{r})$ term.

\subsubsection{Transmission of Gaussian Beams through Cylindrical Lenses}
We compute the properties of a generalized Gaussian beam at the output of a biaxial lens oriented along an arbitrary direction.
The transmittance function of an elliptical lens with focal lengths $f_1$ and $f_2$, oriented at an angle $\theta$ with respect to the $x$ axis, is given by
\begin{equation}
    \label{eq:elliptical_lens_transmittance}
    \mathcal{T}_\theta(f_1, f_2) = \exp\left(-i\dfrac{\pi}{\lambda f_1} (x \cos(\theta) - y \sin(\theta))^2 - i\dfrac{\pi}{\lambda f_2} (x \sin(\theta) + y \cos(\theta))^2 \right).
\end{equation}
The complex electric field of the output beam is computed as the product of the complex electric field of the input beam and the transmittance function. This product effectively maps $\bm{\Lambda}(z)$ to ${\bm{\Lambda}}'(z,\theta)$, where
\begin{equation}
    \label{eq:lambda_elliptical_lens}
    {\bm{\Lambda}}'(z,\theta) = \bm{\Lambda}(z) + \dfrac{i\pi}{\lambda f_1} \begin{bmatrix} \cos^2(\theta) & -\sin(2\theta)/2 \\ -\sin(2\theta)/2 & \sin^2(\theta) \end{bmatrix} + \dfrac{i\pi}{\lambda f_2} \begin{bmatrix} \sin^2(\theta) &  \sin(2\theta)/2 \\ \sin(2\theta)/2 & \cos^2(\theta) \end{bmatrix}.
\end{equation}
This mapping can be used to describe the effect of both spherical and uniaxial cylindrical lenses. For spherical lenses with $f \equiv f_1 = f_2$ that are invariant under rotation, we obtain
\begin{equation}\label{eq:lambda_spherical_lens}
     {\bm{\Lambda}}'(z,\theta) = \bm{\Lambda}(z) + \dfrac{i\pi}{\lambda f} \mathbb{I}_{2 \times 2},
\end{equation}
whereas, for uniaxial cylindrical lenses with $f_2\rightarrow\infty$, we obtain
\begin{equation}\label{eq:lambda_cylindrical_lens}
    {\bm{\Lambda}}'(z,\theta) = \bm{\Lambda}(z) + \dfrac{i\pi}{\lambda f_1} \begin{bmatrix} \cos^2(\theta) & -\sin(2\theta)/2 \\ -\sin(2\theta)/2 & \sin^2(\theta) \end{bmatrix}.
\end{equation}

This mapping can be used to demonstrate the equivalence between a biaxial lens and a pair of uniaxial cylindrical lenses (see Sec.~\ref{sec:solution_biaxial}). Specifically, two cylindrical lenses, each with focal length $f$ and counter-rotated by the angle $\pm\theta$, transform the $\bm{\Lambda}(z)$ matrix in the same way as two cylindrical lenses with focal lengths $f_x = f \sec^2(\theta) / 2$ and $f_y = f \csc^2(\theta) / 2$, aligned along the $x$ and $y$ axes, respectively. This transformation is given by 
\begin{equation}
    {\bm{\Lambda}}'(z,\theta) = \bm{\Lambda}(z) + \dfrac{i\pi}{\lambda \frac{f \sec^2(\theta)}{2}} \begin{bmatrix}
        1 & 0 \\ 0 &  0
    \end{bmatrix} + \dfrac{i \pi}{\lambda \frac{f \csc^2(\theta)}{2}} \begin{bmatrix}
        0 & 0 \\ 0 & 1
    \end{bmatrix}.
\end{equation}

\subsection{Properties of Gaussian Beams transmitted through cylindrical lenses}
\label{sec:gaussian_lens_properties_derivation}

Having established the matrix formalism, we now use it to compute the properties of a Gaussian beam transmitted through a biaxial lens pair. We specifically seek to compute the output beam waists, $z'_{0\mu}$, and beam waist radii, $w'_{0\mu}$, as a function of the input beam properties for $\mu\in \{x,y\}$. We choose the principal axes of the biaxial lens pair to be aligned with the principal axes of the beam chosen to be oriented along the $x$ and $y$ axes. Since the beam propagates independently along both axes, we only compute these quantities for the $x$ axis; the corresponding properties of the beam along the orthogonal $y$ axis can readily obtained.

We consider a diverging beam transmitted through a lens located at $z=0$. The input beam is described by its $\bm{\Lambda}(z)$ matrix,
\begin{equation}
    \bm{\Lambda}(z_{0x}) = \begin{bmatrix} \dfrac{1}{w_{0x}^2} & 0 \\ 0 & * \end{bmatrix},
\end{equation}
where $w_{0x}$ is the beam waist radius measured at its beam waist $z=z_{0x}$.

The $\bm{\Lambda}(z)$ matrix before the lens, $\bm{\Lambda}(0^-)$, is obtained by propagating $\bm{\Lambda}(-z_{0x})$ in free space over a distance $\Delta z=z_{0x}$. Using the transformation rule for free-space propagation, we obtain
\begin{eqnarray}
\bm{\Lambda}(0^-)&=&\left(\bm{\Lambda}(z_{0x})^{-1} - \dfrac{i\lambda z_{0x}}{\pi}\mathbb{I}_{2 \times 2}\right)^{-1}=\begin{bmatrix} w_{0x}^2 - \dfrac{i\lambda z_{0x}}{\pi}  & 0 \\ 0 & * \end{bmatrix}^{-1},
\end{eqnarray}
where $*$ denotes some non-zero arbitrary value. The $\bm{\Lambda}(z)$ matrix after the lens, $\bm{\Lambda}(0^+)$, is obtained by updating $\bm{\Lambda}(0^-)$ using the transformation rule for a uniaxial cylindrical lenses oriented along $\theta=0$,
\begin{eqnarray}
\bm{\Lambda}(0^+)&=&\left(\bm{\Lambda}(0^-) + \dfrac{i\pi}{\lambda f}\right) \begin{bmatrix} 1  & 0 \\ 0 & 0 \end{bmatrix}
=\begin{bmatrix} \left(w_{0x}^2 - \dfrac{i\lambda z_{0x}}{\pi}\right)^{-1}+\dfrac{i\pi}{\lambda f}  & 0 \\ 0 & 0 \end{bmatrix}\label{eq:lambda_a}.
\end{eqnarray}

We seek to compute the beam radii, $w_{0x}'$, and beam waist, $z_{0x}'$ such that 
\begin{eqnarray}
\bm{\Lambda}(z_{0x}')&=&\left(\bm{\Lambda}(0^+)^{-1} + \dfrac{i\lambda z_{0x}'}{\pi}\mathbb{I}_{2 \times 2}\right)^{-1}=\begin{bmatrix} \dfrac{1}{w_{0x}'^2} & 0 \\ 0 & * \end{bmatrix}.
\end{eqnarray}
This equation is equivalent to
\begin{eqnarray}
\bm{\Lambda}(0^+)&=&\left(\bm{\Lambda}(z_{0x}')^{-1} - \dfrac{i\lambda z_{0x}'}{\pi}\mathbb{I}_{2 \times 2}\right)^{-1}
=\begin{bmatrix} w_{0x}'^2 - \dfrac{i\lambda z_{0x}'}{\pi} & 0 \\ 0 & * \end{bmatrix}\label{eq:lambda_b}.
\end{eqnarray}
By comparing Eq.~\eqref{eq:lambda_a} and Eq.~\eqref{eq:lambda_b}, we can solve for the beam waist location
\begin{equation}\label{eq:gaussian_after_lens_z0}
z_{0x}' = \dfrac{\left(1 + \dfrac{z_{0x}}{f}\right)z_{0x} + \left(\dfrac{z_{Rx}}{f}\right)^2 f}{\left(1 + \dfrac{z_{0x}}{f}\right)^2 + \left(\dfrac{z_{Rx}}{f}\right)^2},
\end{equation}
and beam waist radius
\begin{equation}\label{eq:gaussian_after_lens_w0}
w_{0x}' = \dfrac{w_{0x}}{\sqrt{\left(1 + \dfrac{z_{0x}}{f}\right)^2 + \left(\dfrac{z_{Rx}}{f}\right)^2}}.
\end{equation}

Assuming $|(f+z_{0x})z_{0x}| \gg z_{Rx}^2$, we can approximate Eq.~\eqref{eq:gaussian_after_lens_z0} as
\begin{equation}
    z_{0x}' = \dfrac{z_{0x}}{1 + \dfrac{z_{0x}}{f}} + \mathcal{O}\left(\dfrac{z_{Rx}^2}{(f +z_{0x})z_{0x}}\right) \approx \dfrac{f z_{0x}}{f + z_{0x}},
\end{equation}
which can be rearranged into the familiar thin-lens equation from ray optics
\begin{equation}
    \dfrac{1}{z_{0x}'} + \dfrac{1}{-z_{0x}} = \dfrac{1}{f}.
\end{equation}
Similarly, the beam waist radius equation can be obtained from Eq.~\eqref{eq:gaussian_after_lens_w0} by choosing $f = -z_{0x}$,
\begin{equation}
    w_{0x}' = \dfrac{w_{0x}}{z_{Rx} / |f|} = \dfrac{\lambda |f|}{\pi w_{0x}}.
\end{equation}


\section{BEAM CIRCULARIZATION USING A CYLINDRICAL LENS PAIR}\label{sec:derivation_doublet_circularization}

We now derive the conditions on the rotation angle $\theta$ and focal length $f$ required to circularize an astigmatic, elliptical Gaussian beam using two perfect cylindrical lenses. We then show that, even if the lenses are not perfectly uniaxial, there exists an angle $\theta^*$ at which the beam becomes circular.

\subsection{Beam Circularization with Two Perfect Cylindrical Lenses}
If the intensity profile of the beam is not radially symmetric in the plane where the lens pair is placed, then circularization is not possible. The reason is that the lenses only modulate the phase of the beam and do not immediate change its intensity profile. For the beam to become circular after the lens pair, its intensity profile must be radially symmetric at the position of lens pair, i.e., in the plane where the major and minor radii of the Gaussian beam are equal. We therefore assume that the intensity profile of the beam is circular in the plane of the lens pair. We now seek the conditions on the focal length $f$ and rotation angle $\theta$ required to achieve a circular intensity profile in all subsequent planes following the cylindrical lens pair.

We consider an elliptical Gaussian beam whose principal axes are oriented along the $x$ and $y$ axes.
The beam radii along each axis can be described by
\begin{equation}
    \label{eq:elliptical_gaussian_beam_radius}
    w_{\mu}(z) = w_{0\mu} \sqrt{1 + \dfrac{(z-z_{0\mu})^2}{z_{R\mu}^2}} = w_{0\mu} \sqrt{1 + \dfrac{\lambda^2 (z-z_{0\mu})^2}{\pi^2 w_{0\mu}^4}},
\end{equation}
where $w_{0\mu}$ is the beam waist radius at the beam waist $z = z_{0\mu}$ for $\mu\in\{x,y\}$.


We assume that there exists a plane, which we choose as the origin at $z = 0$, where  the intensity profile is radially symmetric, i.e., $r \equiv r_x(0) = r_y(0)$. Then, the $\bm{\Lambda}(z)$ matrix at $z=0$ is given by
\begin{equation}
    \bm{\Lambda}(0) = \begin{bmatrix}
        \dfrac{1}{w_{0x}^2 - \dfrac{j\lambda}{\pi} z_{0x}} & 0 \\
        0 & \dfrac{1}{w_{0y}^2 - \dfrac{j\lambda}{\pi} z_{0y}}.
    \end{bmatrix}
\end{equation}

Given that $r_x(0) = r_y(0)$, and recalling that the eigenvalues of $\Re \bm{\Lambda}(z)$ are given by $r_{\mu}(z)^{-2}$ for $\mu \in \{x, y\}$, we can decompose $\bm{\Lambda}(0)$ into its real and imaginary parts
\begin{equation}
    \bm{\Lambda}(0) = \begin{bmatrix}
        \dfrac{1}{r^2} & 0 \\
        0 & \dfrac{1}{r^2}
    \end{bmatrix} + j\begin{bmatrix}
        \dfrac{\dfrac{\lambda z_{0x}}{\pi}}{w_{0x}^4 + \dfrac{\lambda^2 z_{0x}^2}{\pi^2}} & 0 \\
        0 & \dfrac{\dfrac{\lambda z_{0y}}{\pi}}{w_{0y}^2 + \dfrac{\lambda^2 z_{0y}^2}{\pi^2}}
    \end{bmatrix},
\end{equation}
where
\begin{equation}
    r = r_x(0) = r_y(0) =  w_{0x} \sqrt{1 + \dfrac{z_{0x}^2}{z_{Rx}^2}} = w_{0y} \sqrt{1 + \dfrac{z_{0y}^2}{z_{Ry}^2}}.
\end{equation}
The real part of $\bm{\Lambda}(0)$ has equal eigenvalues, resulting in a radially symmetric intensity profile in the plane $z = 0$. However, the imaginary part of $\bm{\Lambda}(0)$ has unequal eigenvalues, causing the intensity profile to become radially non-symmetric after $z = 0$.

The addition of a cylindrical lens pair at $z = 0$, where both lenses have a focal length $f$ and are counter-rotated by an angle $\pm \theta$, transforms the $\bm{\Lambda}(0)$ matrix to
\begin{equation}
    {\bm{\Lambda}}'(0^+) = \begin{bmatrix} \dfrac{1}{r^2} & 0 \\ 0 & \dfrac{1}{r^2}\end{bmatrix} + j\begin{bmatrix}
        \dfrac{\dfrac{\lambda z_{0x}}{\pi}}{w_{0x}^4 + \dfrac{\lambda^2 z_{0x}^2}{\pi^2}} & 0 \\
        0 & \dfrac{\dfrac{\lambda z_{0y}}{\pi}}{w_{0y}^4 + \dfrac{\lambda^2 z_{0y}^2}{\pi^2}}
    \end{bmatrix} + j\dfrac{2\pi}{f\lambda} \begin{bmatrix}
        \cos^2(\theta) & 0 \\
        0 & \sin^2(\theta)
    \end{bmatrix}.
\end{equation}
If there exists a choice of $f$ and $\theta^*$ such that $\Im {\bm{\Lambda}'}(0^+)$ has two equal eigenvalues, then the beam is circularized. Circularization is thus achieved when
\begin{equation}
    \label{eq:circularization_condition_1}
    \dfrac{\dfrac{\lambda z_{0x}}{\pi}}{w_{0x}^4 + \dfrac{\lambda^2 z_{0x}^2}{\pi^2}} + \dfrac{2\pi}{\lambda f} \cos^2(\theta^*) =  \dfrac{\dfrac{\lambda z_{0y}}{\pi}}{w_{0y}^4 + \dfrac{\lambda^2 z_{0y}^2}{\pi^2}} + \dfrac{2\pi}{\lambda f} \sin^2(\theta^*),
\end{equation}
which can be further simplified to
\begin{equation}
    \label{eq:circularization_condition_2}
    \dfrac{2}{f}\cos^2(\theta^*) - \dfrac{2}{f}\sin^2(\theta^*) = \left(\dfrac{\lambda}{\pi r}\right)^2 \left\{\dfrac{z_{0y}}{w_{0y}^2} - \dfrac{z_{0y}}{w_{0x}^2}\right\}.
\end{equation}
Solving for $\theta^*$, we find the required angle for circularization to be
\begin{equation}
    \label{eq:circularization_angle_proof}
    \theta^* = \dfrac{1}{2} \text{arccos}\left[ \dfrac{f}{2} \left(\dfrac{\lambda}{\pi r}\right)^2 \left\{ \dfrac{z_{0y}}{w_{0w}^2} - \dfrac{z_{0x}}{w_{0x}^2} \right\} \right].
\end{equation}
Since $\theta^*$ is real-valued, the upper bound on $|f|$ for which circularization is possible is given by
\begin{equation}
\label{eq:circularization_focal_length_condition_proof}
    |f|\leq f_{\text{max}} = \bigg\rvert \dfrac{2 \left(\frac{\pi r}{\lambda}\right)^2 }{\frac{z_{0x}}{w_{0x}^2} - \frac{z_{0y}}{w_{0y}^2}} \bigg\rvert.
\end{equation}

We note that this upper bound is always well-defined for a non-circular input Gaussian beam. If the denominator in Eq.~\eqref{eq:circularization_focal_length_condition_proof} vanishes, then the beam is already circular. Therefore, circularization conditions are satisfied for any focal length $f$ provided that $\theta = \pi/4$, i.e., $f_{\text{max}} \to \infty$. The reason is that, at $\theta = \pi/4$, the cylindrical lens pair behaves as a spherical lens, which preserves the circular symmetry of the Gaussian beam.

\subsection{Beam Circularization with Two Imperfect Cylindrical Lenses}\label{sec:circularization_non_uniaxial_lens}

Having found the parameters required to achieve beam circularization using two perfectly uniaxial lenses, we now derive similar conditions for the case where the two cylindrical lenses forming the lens pair are not perfectly uniaxial. We choose each cylindrical lens, denoted as $L_2$ and $L_3$, to have a focal length $f$ along its strong focusing axis, while also possessing a finite, nonzero focal length, $f_{2\perp}$ for $L_2$ and $f_{3\perp}$ for $L_3$, along the weak focusing axis, which is perpendicular to the strong axis.. 


\subsubsection{Two Identical Lenses}
In the symmetric case where $f_{2\perp} = f_{3\perp} = f_{\perp}$, each of the two imperfect cylindrical lenses can be modeled as two perfectly-uniaxial cylindrical lenses. The four equivalent lenses can then be paired into two effective biaxial lenses. The first two equivalent lenses, $L_{2,1}$ and $L_{3,1}$, are aligned with the strong focusing axis of $L_2$ and $L_3$, respectively. Together, both lenses form an effective biaxial lens pair comprising two uniaxial lenses with focal length $f$ and a relative angle of $2\theta$. The two other equivalent lenses, $L_{2,2}$ and $L_{3,2}$, are oriented perpendicular to the weak focusing axis of $L_2$ and $L_3$, respectively, with a focal length $f_{\perp}$. Together, both lenses form a biaxial lens pair with focal length of $f_\perp$, each counter-rotated by an angle $\pi/2+\theta$.

Recalling that two uniaxial cylindrical lenses with focal length $f$ mounted one behind the other at a relative angle $2\theta$ formed an effective biaxial lens with (see Eq.~\eqref{eq:transmittance_equivalence3}), we find
\begin{eqnarray}
f_x^A = \frac{f}{2}\sec^2(\theta)\\
f_y^A = \frac{f}{2}\sec^2(\theta)
\end{eqnarray}
for the first effective biaxial lens, and 
\begin{eqnarray}
f_x^B(\theta) &=& \frac{f_\perp}{2}\sec^2(\pi/2+\theta)=\frac{f_\perp}{2}\csc^2(\theta)\\
f_y^B(\theta) &=& \frac{f_\perp}{2}\csc^2(\pi/2+\theta)=\frac{f_\perp}{2}\sec^2(\theta)
\end{eqnarray}
for the second effective biaxial lens.
Then, considering that two cylindrical lenses, both oriented in the same $\mu$ direction, with focal lengths $f_\mu^A$ and $f_\mu^B$, act as a single cylindrical lens with focal length
\begin{equation}
    \frac{1}{f_\mu} = \dfrac{1}{f_\mu^{A}} + \dfrac{1}{f_\mu^B},
\end{equation} 
we find
\begin{eqnarray}
    \dfrac{1}{f_x(\theta)} &=& 2 \left(\dfrac{\cos^2(\theta)}{f} + \dfrac{\sin^2(\theta)}{f_{\perp}}\right)\\
    \dfrac{1}{f_y(\theta)} &=& 2 
    \left(\dfrac{\sin^2(\theta)}{f} + \dfrac{\cos^2(\theta)}{f_{\perp}}\right)
\end{eqnarray}
for the effective focal lengths of the cylindrical lens pair. The circularization condition for this pair is almost identical to the one derived in Eq.~\eqref{eq:circularization_condition_2},
\begin{equation}
    \dfrac{1}{f_{x}(\theta^*)} - \frac{1}{f_{y}(\theta^*)} = \left(\dfrac{\lambda}{\pi r}\right)^2 \left\{\dfrac{z_{0y}}{w_{0y}^2} - \dfrac{z_{0x}}{w_{0x}^2}\right\},
\end{equation}
from which we find
\begin{equation}
    \label{eq:circularization_angle_proof_biaxials}
    \theta^* = \dfrac{1}{2} \text{arccos}\left[ \dfrac{{f}'}{2} \left(\dfrac{\lambda}{\pi r}\right)^2 \left\{ \dfrac{z_{0yy}}{w_{0y}^2} - \dfrac{z_{0x}}{w_{0x}^2} \right\} \right],
\end{equation}
where
\begin{equation}
    \label{eq:equivalent_lens}
    {f}' = \left(\dfrac{1}{f} - \dfrac{1}{f_{\perp}}\right)^{-1}.
\end{equation}
The condition for the existence of an angle $\theta^*$ for which circularization is possible is given by 
\begin{equation}
    |{f}'| \leq f_\text{max}=\bigg\rvert \dfrac{2 \left(\frac{\pi r}{\lambda}\right)^2 }{\frac{z_{0y}}{w_{0y}^2} - \frac{z_{0x}}{w_{0x}^2}} \bigg\rvert.
\end{equation}

These results are identical to those derived for a perfectly uniaxial cylindrical lens pair (see Eqs.~\eqref{eq:circularization_angle_proof}--\eqref{eq:circularization_focal_length_condition_proof}), except that the nominal focal length $f$ is replaced by the equivalent focal length $f'$.

\subsubsection{Two Non-Identical Lenses}
In the asymmetric case where $f_{2\perp} \neq f_{3\perp}$, perfect circularization is no longer achievable, because for any angle $\theta$, the lens pair introduces a residual phase curvature to the beam that makes the beam elliptical and astigmatic. This statement can be proved mathematically. Suppose $f_{3\perp} = f_{2\perp} + \delta$, where $\delta \neq 0$. The lens with focal length $f_{2\perp}$ is equivalent to a pair of two uniaxial cylindrical lenses with focal lengths $f_{3\perp}$ and $f_{\delta}$, where
\begin{equation}
    \label{eq:fdelta_definition}
    f_{\delta} = \left( \dfrac{1}{f_{2\perp}} - \dfrac{1}{f_{3\perp}} \right)^{-1} = f_{2\perp} \left(1 - \dfrac{1}{1 + \delta / f_{2\perp}} \right)^{-1}.
\end{equation}
This result is a direct consequence of the fact that two cylindrical lenses with focal lengths $f_1$ and $f_2$ mounted one behind the other along the same orientation are equivalent to a uniaxial cylindrical lens with focal length
\begin{equation}
    f = \left(\dfrac{1}{f_1} + \dfrac{1}{f_2}\right)^{-1}.
\end{equation}

An important consequence of Eq.~\eqref{eq:fdelta_definition} is that when $f_{2\perp}$ and $f_{3\perp}$ have the same sign, then $|f_{\delta}| > \min(|f_{2\perp}|, |f_{3\perp}|)$. This statement indicates that the equivalent lens with focal length $f_{\delta}$ introduces less focusing power than the more defective of the two cylindrical lenses in the direction perpendicular to their nominal focusing axes.

If the effect of $f_{\delta}$ is neglected, then the system effectively reduces to a lens pair with $f_{2\perp} = f_{3\perp}$, for which perfect circularization is possible. Therefore, the lens pair first circularizes the input beam and subsequently acts on the circularized beam with a uniaxial cylindrical lens of focal length $f_{\delta}$, oriented at an angle $\theta^* + \pi/2$ with respect to the $x$ axis. The resulting beam is an elliptical Gaussian beam with astigmatism oriented at $\theta^* + \pi/2$ relative to the $x$ axis.

Note that this effect is expected to be negligible in practice for many tabletop optics applications. For instance, from our experimental measurements (see Sec.~\ref{sec:experiment} in the main text), we estimate a transverse focal length of $f_{\perp} = 51.4~\text{m}$, where $f_{2\perp}$ and $f_{3\perp}$ are equal to within measurement precision, i.e. $\delta \ll f_{2\perp}$. These results imply
\begin{equation}
    f_{\delta} = \dfrac{f_{2\perp}^2}{\delta} + \mathcal{O}(\delta^2) \gg f_{2\perp},
\end{equation} 
making the residual astigmatism and ellipticity caused by the difference between the lenses negligible.

\section{ASTIGMATISM CORRECTION USING A CYLINDRICAL LENS PAIR}
\label{sec:derivation_doublet_deastigmatization}

We now prove that under appropriate conditions, a pair of cylindrical lenses with focal lengths $f$ can transform an elliptical Gaussian beam with astigmatism into an elliptical Gaussian beam without astigmatism. 
We then propose a method to compute the set of viable focal lengths for which correcting astigmatism is possible.


\subsection{Proof of the Existence of an Angle for Astigmatism Correction}

We first show that, for sufficiently small values of $|f|$, a cylindrical lens pair can correct astigmatism for any elliptical Gaussian beam. We assume that the lens pair is located in the plane $z=0$. The input beam has a beam waist radii of $w_{0x}$ and $w_{0y}$ at the beam waists located at $z=z_{0x}$ and $z=z_{0y}$, respectively, where $z_{0x},~z_{0y}$ may be positive or negative. After passing through the lens pair, the output beam has beam waist radii $w_{0x}'$ and $w_{0y}'$ at locations $z_{0x}'$ and $z_{0y}'$, respectively (see Eqs.~\eqref{eq:gaussian_after_lens_z0}--\eqref{eq:gaussian_after_lens_w0} in Sec.~\ref{sec:gaussian_lens_properties_derivation} for explicit definitions derived in terms of the input beam parameters). The lens pair acts as an effective biaxial lens with focal lengths $f_x(\theta) = f \sec^2(\theta)/2$ and $f_y(\theta) = f \csc^2(\theta) / 2$ along the $x$ and $y$ axes, respectively. We define the distance between the beam waist locations
\begin{equation}
    \Delta z(\theta) = z_{0y}'(\theta) - z_{0x}'(\theta).
\end{equation}
Correcting astigmatism requires showing that there exists values of $f$ and $\theta$ for which the beam waists overlap, i.e.,  $\Delta(z)(\theta) = 0$. We solve this problem for three possible cases that cover all possible configurations of $z_{0x}$ and $z_{0y}$. 

\subsubsection{Case 1: $z_{0x} \leq 0$ and $z_{0y} \leq 0$}

We prove that, for sufficiently small and positive values of $f$, there always exists at least one angle $\theta^*$ such that $\Delta(z)(\theta^*) = 0$. We choose $f$ such that $0 < f \ll 1$. When $\theta = 0$, it follows from Eq.~\eqref{eq:gaussian_after_lens_z0} that $z_{0x}'(0) = f/2 + \mathcal{O}(f^2)$ and $z_{0y}'(0) = z_{0y}$. For small enough values of $f$ where the quadratic term is smaller than $f/2$, we have $z_{0x}'(0) > 0$ and $z_{0y}'(0) \leq 0$.

Similarly, when $\theta = \pi/2$, we have $z_{0x}'\left(\pi/2\right) = z_{0x}$, and $z_{0y}'\left(\pi/2\right) = f/2 + \mathcal{O}(f^2)$, such that, for small values of $f$, we have $z_{0x}'\left(\dfrac{\pi}{2}\right) \leq 0$ and $z_{0y}'\left(\dfrac{\pi}{2}\right) > 0$.
As a result of these inequalities, we have $\Delta z(0) < 0$ and $\Delta z\left(\dfrac{\pi}{2}\right) > 0$, so that by the Intermediate Value Theorem for continuous functions, there exists at least one angle $\theta \in \left[0, \pi/2\right]$ such that $\Delta z(\theta) = 0$.


\subsubsection{Case 2: $z_{0x} \geq 0$ and $z_{0y} \geq 0$}

The proof for Case 2 is similar to Case 1. We choose $f$ such that $f < 0$ and close to $0$. We can then show that $\Delta z(0) > 0$ and $\Delta z\left(\pi/2\right) < 0$ for sufficiently small values of $|f|$. Using the Intermediate Value Theorem, we can show that there exists some $\theta \in \left[0, \pi/2\right]$ for which $\Delta z(\theta) = 0$.

\subsubsection{Case 3: $z_{0x} \cdot z_{0y} < 0$}
In this case, the cylindrical lens pair is placed in between the planes $z = z_{0x}$ and $z = z_{0y}$, making the proof slightly different from the previous two cases. We choose $f$ such that $f < 0$ and close to $0$. Without loss of generality, assuming that $z_{0x} < 0$ and $z_{0y} > 0$, we find $z_{0x}'\left(\pi/2\right) = z_{0x}$, and $z_{0y}'\left(\pi/2\right) = f/2 + \mathcal{O}(f^2)$. Therefore, for sufficiently small values of $|f|$
\begin{equation}
    \label{eq:desatigmatism_proof_evidence_5}
    \Delta z(\pi/2) = z_{0y}'(\pi/2) - z_{0x}'(\pi/2) > 0.
\end{equation}

Now we prove that there exists some angle $\theta$ close to $0$ such that $\Delta z(\theta) < 0$ when choosing a small enough value of $|f|$. We choose $\phi$ such that $f_y(\phi) = f \csc^2(\phi) / 2 = -z_{0y}$, when 
\begin{equation}
\phi = \sin^{-1}\left(\sqrt{\frac{-f}{2z_{0y}}}\right).
\end{equation}

Since we have chosen $f$ to satisfy the condition $-1 \ll f < 0$, $\phi$ can be made arbitrarily close to $0$. As a result, $f_x(\phi)$ will be close to $f_x(0) = f/2$. More precisely
\begin{equation}
    f_x(\phi) = \dfrac{f}{2} \sec^2(\phi) = \dfrac{f}{2} + \mathcal{O}(f^2).
\end{equation}
For this choice of $\phi$ and following Eq.~\eqref{eq:gaussian_after_lens_z0}, we conclude that $z_{0x}'(\phi) = \dfrac{f}{2} + \mathcal{O}(f^2)$, and $z_{0y}'(\phi) = -z_{0y}$. Therefore, $z_{0x}'(\phi) > 0$, and $z_{0y}'(\phi) < 0$. As a result of these inequalities we have $\Delta z(\phi) < 0$.
From these inequalities and the Intermediate Value Theorem, we conclude that there exists some $\theta \in \left[\phi, \pi/2\right] \subset \left[0, \pi/2\right]$ where $\Delta z(\theta) = 0$.
 
\subsection{Computing the Bounds on the Lens Pair Focal Length for De-Astigmatization}

We seek the necessary conditions on the focal length $f$ to correct astigmatism in elliptical Gaussian beams. Our approach involves finding the optimal angle $\theta^*(f)$ for different values of $f$. The domain of this function corresponds to the range of valid choices of $f$ that allow for correcting astigmatism. It follows from Eq.~\eqref{eq:gaussian_after_lens_z0} that given an angle $\theta$, a necessary condition to correct astigmatism in an elliptical Gaussian beam is
\begin{equation}
\label{eq:deastigmatization_full_form}
    z_{0x}'(f_x) = \dfrac{z_{0x} f_x(\theta)^2 + (z_{0x}^2 + z_{Rx}^2) f_x(\theta)}{(z_{0x} + f_x(\theta))^2 + z_{Rx}^2} = \dfrac{z_{0y} f_y(\theta)^2 + (z_{0y}^2 + z_{Ry}^2) f_y(\theta)}{(z_{0y} + f_y(\theta))^2 + z_{Ry}^2} = z_{0y}'(f_y(\theta)),
\end{equation}
where $f_x = f_x(\theta) = f \sec^2(\theta) / 2$ and $f_y = f_y(\theta) = f \csc^2(\theta) / 2$. Solving this equation for $\theta$ results in a complicated equation involving multiple powers of $\sin(\theta)$ and $\cos(\theta)$ without a simple representation.


Alternatively, we can fix $f_y$ and search for the $f_x$ required to the correct astigmatism. Having both $f_y$ and $f_x(f_y)$, we can infer the required angle $\theta$ and focal length $f$ of the lens pair. Solving for $f_x$ in Eq.~\eqref{eq:deastigmatization_full_form}, we find that
\begin{equation}
    f_x(f_y) = \dfrac{A(f_y) \pm \sqrt{ A(f_y)^2 - 4 z_{0y}'(f_y) (z_{0y}'(f_y) - z_{0x})( z_{0x}^2 + z_{Rx}^2 ) }}{2\left(z_{0y}'(f_y) - z_{0x}\right)},
\end{equation}
where $A(f_y) = z_{0x}^2 + z_{Rx}^2 - 2 z_{0y}'(f_y) z_{0x}$. For every $f_y$, we find the values of $f$ and $\theta$ that correct astigmatism, namely
\begin{eqnarray}
    \label{eq:deastigmatization_equation_1}
    \dfrac{2}{f(f_y)} &=& \dfrac{1}{f_y} + \dfrac{1}{f_x(f_y)},
\end{eqnarray}
and
\begin{eqnarray}
    \label{eq:deastigmatization_equation_2}
    \theta(f_y) &=& \arctan\left( \sqrt{ \dfrac{f_x(f_y)}{f_y} } \right).
\end{eqnarray}
These equations produce valid choices of $f$ and $\theta$ as long as the conditions $|f_y| > \dfrac{|f|}{2}$ and $f \cdot f(f_y) > 0$ are satisfied. These conditions guarantee that $f_x(f_y)$ and $f_y$ have the same sign and are both larger than $f/2$. Consequently, there always exists a valid choice of $f$ and $\theta$ that can be used to correct astigmatism in elliptical Gaussian beams.


\section{ELLIPTICITY AND ASTIGMATISM CORRECTION WITH THREE CYLINDRICAL LENSES}\label{sec:first_lens_existence}

In this section, we demonstrate that a cylindrical lens oriented along the major axis of an elliptical Gaussian beam, with an appropriate choice of focal length, guarantees the existence of a plane after the lens where the intensity profile becomes radially symmetric. We first prove that such a plane exists, provided the lens is aligned precisely along the major axis. We then study the resulting intensity profile when the orientation of the cylindrical lens deviates from the major axis.

\subsection{Circularization with Perfectly Oriented Cylindrical Lenses}

To circularize an elliptical Gaussian beam with astigmatism using a pair of cylindrical lenses, the intensity profile of the beam must be made radially symmetric in the plane where the lens pair is located. The first cylindrical lens in our three-lens solution, placed in the plane $P_1$ at $z = z_{P_1}$, fulfills this role by generating a radially symmetric intensity profile in a plane $P_2$ at $z = z_{P_2}$, where the lens pair is located (see Sec.~\ref{sec:solution_circularization}).

We now prove that, for any input beam, there always exists a suitable choice of focal length $f_1$ such that the plane $P_2$ exists. We assume that at $z = z_{P_1}$, the radius of the beam along the $y$ axis is larger than the radius along the $x$ axis, i.e., $w_{0y}({z_{P_1}})>w_{0x}({z_{P_1}})$, such that $\Delta w_{xy}(z_{P_1}) = w_y(z_{P_1}) - w_x(z_{P_1})>0$. We choose the focal length of the first lens, oriented along $y$, to be $f_1 > 0$ and $f_1 \ll z_{R\mu}$ for both $\mu \in \{x, y\}$. 

Using the transformation properties of Gaussian beams under lens action (see App.~\ref{sec:gaussian_lens_properties_derivation}), we can show that the beam radius along the $y$ axis focuses at
\begin{equation}
z_{0y} = z_{P_1} + f_1 + \mathcal{O}(f_1^2),
\end{equation}
with a beam waist radius of
\begin{equation}
w_{0y} = \mathcal{O}(f_1).
\end{equation}
For sufficiently small $f_1$, we have $w_{0y} < w_{0x}$ and $z_{0y} > z_{P_1}$. The beam focuses along the $y$ axis after $z = z_{P_1}$ and achieves a smaller waist than along the $x$ axis. As a result, the beam radius difference becomes negative at $z = z_{0y}$, i.e., $\Delta w(z_{0y}) = w_y(z_{0y}) - w_x(z_{0y}) < 0.$
Since $\Delta w(z)$ is a continuous function of $z$, and $\Delta w(z_{P_1}) > 0$ whereas $\Delta w(z_{0y}) < 0$, it follows from the Intermediate Value Theorem that there exists some $z_{P_2} \in [z_{P_1}, z_{0y}]$ such that $\Delta w(z_{P_2}) = 0$. The plane $P_2$ is the plane at which the beam achieves a radially symmetric intensity profile, i.e., $w_x(z_{P_2}) = w_y(z_{P_2})$. Hence, for any elliptical Gaussian beam, we can always choose a sufficiently small focal length $f_1$ such that a symmetry plane $z_{P_2}$ exists.

\subsection{Limitations of Imperfectly Oriented Cylindrical Lenses}\label{sec:imperfect_orientation_anti_crossing}

\begin{figure}[t!]
\centering
\includegraphics[]{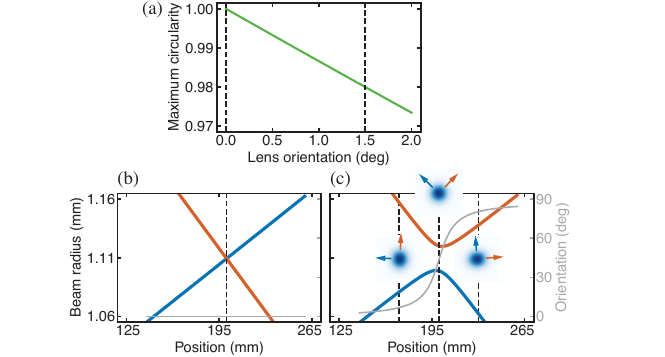}
\caption{
\label{fig:crossing}
\textbf{Imperfect lens orientation.}
A cylindrical lens of focal length $f=500~\text{mm}$ is placed at $z=0~\text{mm}$ at an angle $\theta$ with respect to the major axis of an incoming elliptical beam. The maximum circularity of the beam decreases away from $1.0$ when the cylindrical lens is rotated away from $\theta = 0^{\circ}$. At $\theta = 0^{\circ}$ the beam radii intersect at $z=200~\text{mm}$ and the beam orientation remains constant. At $\theta = 1.5^{\circ}$ the beam radii of the beam do not intersect, and the beam rotates by almost $90^{\circ}$. The beam reaches a maximum circularity of $0.98$ at $z=200~\text{mm}$.
}
\end{figure}

We have so far assumed that the first cylindrical lens is perfectly aligned with the major axis of the elliptical beam. In that case, given an appropriate choice of focal length $f_1$, there exists a plane $P_2$ at which the beam acquires a circular intensity profile. We now show that, if the lens is rotated with respect to the the elliptical beam, then perfect circularization cannot be realized. We then analyze how the circularity of the beam in the plane $P_2$ degrades as a function of the misalignment angle (see Fig.~\ref{fig:crossing}).

Let the location of the first cylindrical lens to be $z = 0$. When the lens is not properly aligned, the output is a generalized Gaussian beam with a $\bm{\Lambda}$ matrix whose real and imaginary parts do not commute: $[\Re \bm{\Lambda}(0^+), \Im \bm{\Lambda}(0^+)] \neq 0$. This nonzero commutator implies that the beam remains asymmetric at all planes beyond $z > 0$. Since the beam propagates in free space for $z > 0$, the commutator $[\Re \bm{\Lambda}(z), \Im \bm{\Lambda}(z)]$ remains nonzero at all such positions due to the transformation properties of the $\bm{\Lambda}$ matrix (see Table~\ref{table:lambda_rules}). However, for the beam to exhibit a radially symmetric intensity profile at some plane $z = z_0$, the commutator at that plane must vanish. This contradiction implies that no such plane can exist.

To clarify this point, recall that for a generalized Gaussian beam (see Eq.~\eqref{eq:gaussian_generalized}), the real part of $\bm{\Lambda}(z)$ can be expressed in terms of the beam radii $r_\mu$ with $\mu \in \{0, 1\}$ and the beam orientation angle $\theta$:
\begin{equation}
    \Re\left\{\bm{\Lambda}(z)\right\} = 
    \begin{bmatrix}
        \cos \theta & -\sin \theta \\
        \sin \theta & \cos \theta
    \end{bmatrix}
    \begin{bmatrix}
        \frac{1}{r_0^2} & 0 \\
        0 & \frac{1}{r_1^2}
    \end{bmatrix}
    \begin{bmatrix}
        \cos \theta & \sin \theta \\
        -\sin \theta & \cos \theta
    \end{bmatrix}.
\end{equation}
If the intensity profile of the beam is radially symmetric, the beam radii must be equal, $r = r_0 = r_1$, simplifying the previous expression to
\begin{equation}
    \Re\bm{\Lambda}(z) = \frac{1}{r^2} \mathbb{I}_2.
\end{equation}
Since the $\mathbb{I}_2$ identity matrix commutes with any other matrix, we have $[\Re \bm{\Lambda}, \Im \bm{\Lambda}] = 0$. Hence, the beam is a Gaussian beam. This derivation proves that achieving a radially symmetric intensity profile is only possible if the beam is not a generalized Gaussian beam. 

We now study through numerical simulations how misalignment of the first cylindrical lens affects the maximum circularity of the beam. 
We consider a perfectly uniaxial cylindrical lens with a focal length of $f=500~\text{mm}$. We find
a $2^{\circ}$ misalignment of between the focusing axis of the lens and the major axis of the beam and the reduces the maximum achievable circularity from $1.0$ when perfectly aligned to $0.97$ (see Fig.~\ref{fig:crossing}). 

We further examine how the beam profile obtained for $\theta = 1.5^{\circ}$ compares against the one obtained for $\theta = 0^{\circ}$ (see insets in Fig.~\ref{fig:crossing}). When $\theta = 0^{\circ}$, the beam orientation remains fixed at $0^{\circ}$, and the beam radii intersect at $z = 200~\text{mm}$. In contrast, when $\theta = 1.5^{\circ}$, the beam orientation undergoes a rotation of nearly $90^{\circ}$, and the radii never intersect. As a result, the circularity peaks at $0.98$ near $z = 200~\text{mm}$, placing a practical limit on the degree of circularization achievable using our three-lens solution.

\section{IMAGE PROCESSING METHOD FOR EXTRACTING BEAM PROPERTIES}
\label{sec:image_analysis}

\begin{figure}[t!]
\centering
\includegraphics[]{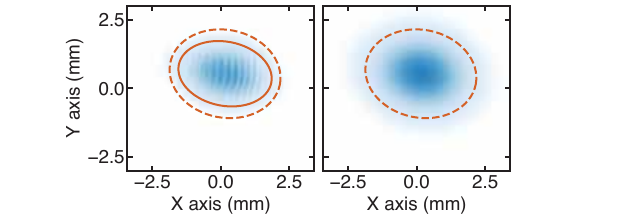}
\caption{
\label{fig:image_analysis}
\textbf{Image processing.}
(left)~The raw image captured by the camera exhibits interference effects caused by its protective glass cover. The beam radii along the principal axes are extracted from a rescaled best-fit ellipse (solid red) applied to the filtered image (dashed red).
(right)~The filtered image is obtained by applying a Gaussian filter with $\sigma = 0.55~\text{mm}$ to the raw image. The principal axes are extracted by fitting the filtered image to the best-fit ellipse (dashed red).
}
\end{figure}

We now describe the image processing algorithm used to extract properties of the beam from intensity profiles captured at various positions along the propagation axis. From each image, we determine the orientation of the beam and extract the beam radii along the principal axes. For the analysis, we assume that the beam can be modeled by a generalized Gaussian beam (see App.~\ref{sec:matrix_formalism}) whose intensity profile is given by Eq.~\eqref{eq:generalized_gaussian_intensity}. Given a measurement of the intensity profile on a camera, the image processing task consists of extracting and diagonalizing the $2\times 2$ real-valued and symmetric matrix $\Re{\bm{\Lambda}}(z)$.

\subsection{Filtering of Interference Patterns}
We first describe the method used to digitally filter the interference patterns observed when imaging the intensity profile of the beam on a CMOS camera (Basler ace2 a2A2840-48umBAS). 
The sensor of this camera has a protective glass cover that results in multiple reflections of the beam between the sensor and the glass cover. These reflections result in Fabry–Pérot fringes on the image (see Fig.~\ref{fig:image_analysis}a), which increase the complexity of the image processing algorithm and can lead to inaccurate estimates of the beam properties.

To remove these interference patterns, we first apply a Gaussian blur using a Gaussian kernel with a suitably chosen standard deviation, $\sigma$. The Gaussian blur acts as a two-dimensional low-pass filter, suppressing high-frequency interference fringes while slightly broadening the apparent size of the beam. The apparent increase in the beam size can be quantified using the properties of Gaussian functions. The convolution of two Gaussian functions with zero mean and standard deviations $\sigma_1$ and $\sigma_2$ results in another Gaussian function with zero mean and standard deviation $\sqrt{\sigma_1^2 + \sigma_2^2}$. The radius $r$ of a generalized Gaussian beam is related to the standard deviation $\sigma$ of the Gaussian function describing its intensity profile by the relation $r = 2 \sigma$. Therefore, given the standard deviation $\sigma$ of the Gaussian kernel used in the blur and the apparent radius of the filtered image $r_{\text{apparent}}$, the true radius of the beam $r_{\text{true}}$ is given by
\begin{equation}
\label{eq:gaussian_blur_correction}
    r_{\text{true}} = \sqrt{r_{\text{apparent}}^2 - 4\sigma^2}.
\end{equation}

\subsection{Extraction of Beam Parameters}
We now describe the method used for extracting the orientation and radii of the beam from the filtered images. We denote the measured pixel values at the $i$-th row and $j$-th column of the camera sensor by $p_{ij} \in [0, 255]$. We first apply a binary mask to the filtered image, setting all pixels with values less than $\max_{ij} p_{ij} / e^{2}$ to zero (see the insets of Fig.~\ref{fig:image_analysis}). We then extract the coordinates of the pixels forming the boundary of the island of non-zero pixels remaining after the binary mask is applied using the \textit{findContours} function of the OpenCv2 image analysis library~\cite{OpenCv}. We finally extract the ellipse of best fit to the pixel coordinates of the boundary using the \textit{fitEllipse} function of OpenCv2, which uses a direct least square method~\cite{Fitzgibbon1995}. This fit returns the orientation and radii of the best-fit ellipse, from which the beam orientation and radii are obtained after applying a Gaussian blur correction. 

\end{appendix}


\end{document}